\documentclass[a4paper,11pt]{article}
\pdfoutput=1 % if your are submitting a pdflatex (i.e. if you have
\usepackage{lineno}
\usepackage{todonotes}
\usepackage{siunitx}
\usepackage{comment}
\usepackage{multirow}
\usepackage[normalem]{ulem}
\usepackage{jinstpub} % for details on the use of the package, please
\title{\boldmath Accurate modelling of 3D-trench silicon sensor with  enhanced timing performance and comparison with test beam measurements}

\author[a]{D.~Brundu,}
\author[a]{A.~Cardini,}
\author[a]{A.~Contu,}
\author[a]{G.~M.~Cossu,}
\author[b,e]{G.-F.~Dalla Betta,}
\author[a,f]{M.~Garau,}
\author[a]{A.~Lai,}
\author[a,f]{A.~Lampis,}
\author[a]{A.~Loi,}
\author[c,g]{M.~M.~Obertino,}
\author[d]{B.~G.~Siddi,}
\author[d]{S.~Vecchi}
\affiliation[a]{INFN, Sezione di Cagliari, Cagliari, Italy}
\affiliation[b]{INFN, TIFPA, Trento, Italy}
\affiliation[c]{INFN, Sezione di Torino, Torino, Italy}
\affiliation[d]{INFN, Sezione di Ferrara, Ferrara, Italy}
\affiliation[e]{Dipartimento di Ingegneria Industriale, Universit\`a di Trento, Trento, Italy}
\affiliation[f]{Dipartimento di Fisica dell'Universit\`a di Cagliari, Cagliari, Italy}
\affiliation[g]{Dipartimento di Scienze Agrarie, Forestali ed Alimentari dell'Universit\`a di Torino, Grugliasco, Italy}

\abstract{

This paper presents the detailed simulation of a double-pixel structure for charged particle detection based on the 3D-trench silicon sensor developed for the TIMESPOT project
and a comparison of the simulation results with measurements performed at $\pi-$M1 beam at PSI laboratory.
The simulation is based on the combined use of several software tools (TCAD, GEANT4, TCoDe and TFBoost) which allow to fully design and simulate the device physics response in very short computational time, ${\cal O}(1-100{\rm s})$ per simulated signal, by exploiting parallel computation using single or multi-thread processors.
This allowed to produce large samples of simulated signals, perform detailed studies of the sensor characteristics and make precise comparisons with experimental results.}


\begin{document}
\maketitle
\flushbottom

\section{Introduction}\label{sec:intro}

The harsh conditions foreseen at the next generation of collider experiments (including the LHC Run-5 upgrades) require vertex and tracking detectors with very high radiation resistance and high-resolution timing capabilities~\cite{CMS-TL,ATLAS-TL,LHCb-PII-Physics,FCC}. Silicon sensors with 3D structure appear as a very suitable technological solution in this respect. Owing to their specific layout, they show unmatched resistance to particle fluence~\cite{3Dradhard} and are intrinsically fast. Moreover, the specific electrode structure, perpendicular to the particle-impinging surface, allows large flexibility in studying and designing a pixel geometry optimised for best timing performance.

The TIMESPOT project carries out a development program to design, produce and characterise 3D pixel sensors with enhanced timing capabilities, leading to resolutions which have been already measured below 30~ps under test beam~\cite{Hiroshima, JINST-TimeSpot}. The present paper illustrates the simulation and design activities which have lead to such an excellent experimental result.
	
\section{Static modelling of 3D silicon sensors}
Precise timing resolution requires fast signals and high signal-to-noise ratio. These requirements have precise implications on the performance of the sensor and of the front-end electronics. On the sensor side, it is crucial to have uniform signal shapes and a narrow distribution of charge collection times at the electrodes. In the case of 3D devices this feature can be obtained with a suitable design of the pixel geometry. According to the Ramo theorem~\cite{Ramo}, to obtain induced current signals with uniform shapes, it is essential to have uniform values of the weighting field and of the charge carriers drift velocities across the sensible volume of the pixel. This can be achieved by a parallel trench-based square pixel layout as shown in this paper. Compared to other 3D geometries (based on different hexagonal and square pixel shapes with various configurations of trenched and columnar electrodes) this layout shows the best properties, as detailed in ref.~\cite{TCode}. 

\subsection{Geometry definition}
\begin{figure}[b!]
	\centering
		\includegraphics[width=0.8\linewidth]{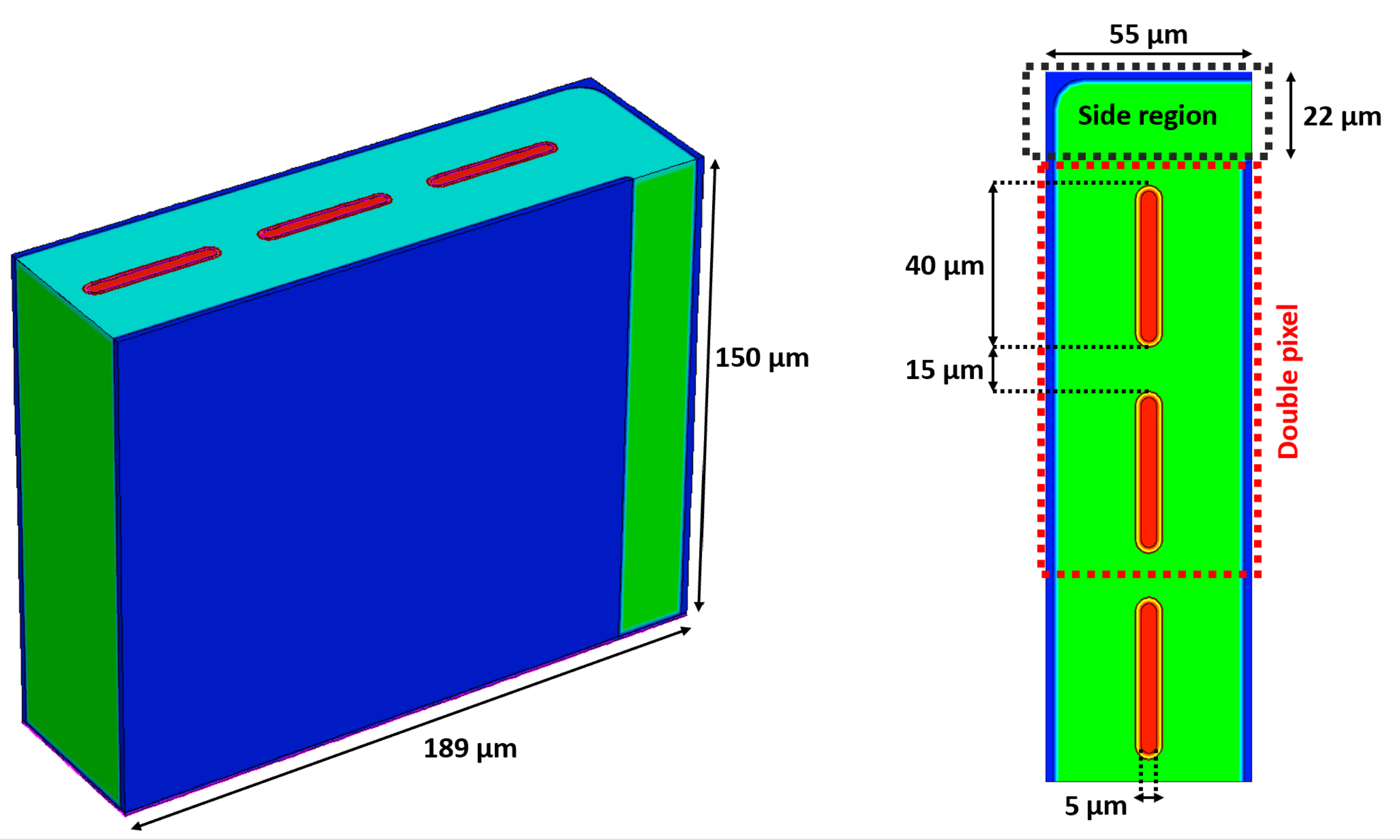}
		\caption{Layout of the simulated TIMESPOT test structure, including sections and sizes, designed using Sentaurus TCAD. The double pixel is indicated by the dotted-red line.}
		\label{fig:TestStructure}
	\end{figure}
The specific geometry considered in the present paper for modellisation and analysis is based on a parallel electrode configuration made up of three trenches (figure~\ref{fig:TestStructure}). The pixel is fabricated on a \SI{150}{\micro\meter} thick, high-resistivity silicon wafer which is wafer-bonded~\cite{directBond} on a secondary, high-conductivity wafer used as mechanical support during the fabrication and providing the bias voltage for the resistive electrodes. The pixel features a pitch of \SI{55}{\micro\meter}. Two ohmic trenches (represented in blue in figure~\ref{fig:TestStructure}) of dimensions 2.5x55x150 \SI{}{\micro\meter}$^3$ are located at the two opposite sides of the pixel. A third trench of dimensions 5x40x130 \SI{}{\micro\meter}$^3$ is placed at the pixel centre, parallel to the two ohmic electrodes and serves as readout electrode (represented in red in figure~\ref{fig:TestStructure}). The model designed for the simulation is an exact replica of the double-pixel device tested at PSI in 2019~\cite{JINST-TimeSpot}. It consists of two standard parallel trench pixels connected to the same readout electrode (referred to as double pixel) and a third neighbouring pixel, connected to ground, to better describe the boundary conditions of the active pixels (electric and weighting fields). The double pixel is located at the border of a test structure as shown in figure~\ref{fig:TestStructure}.

\subsection{Calculation of the field maps}
\begin{figure}[b!]
	\centering
		\includegraphics[width=0.95\linewidth]{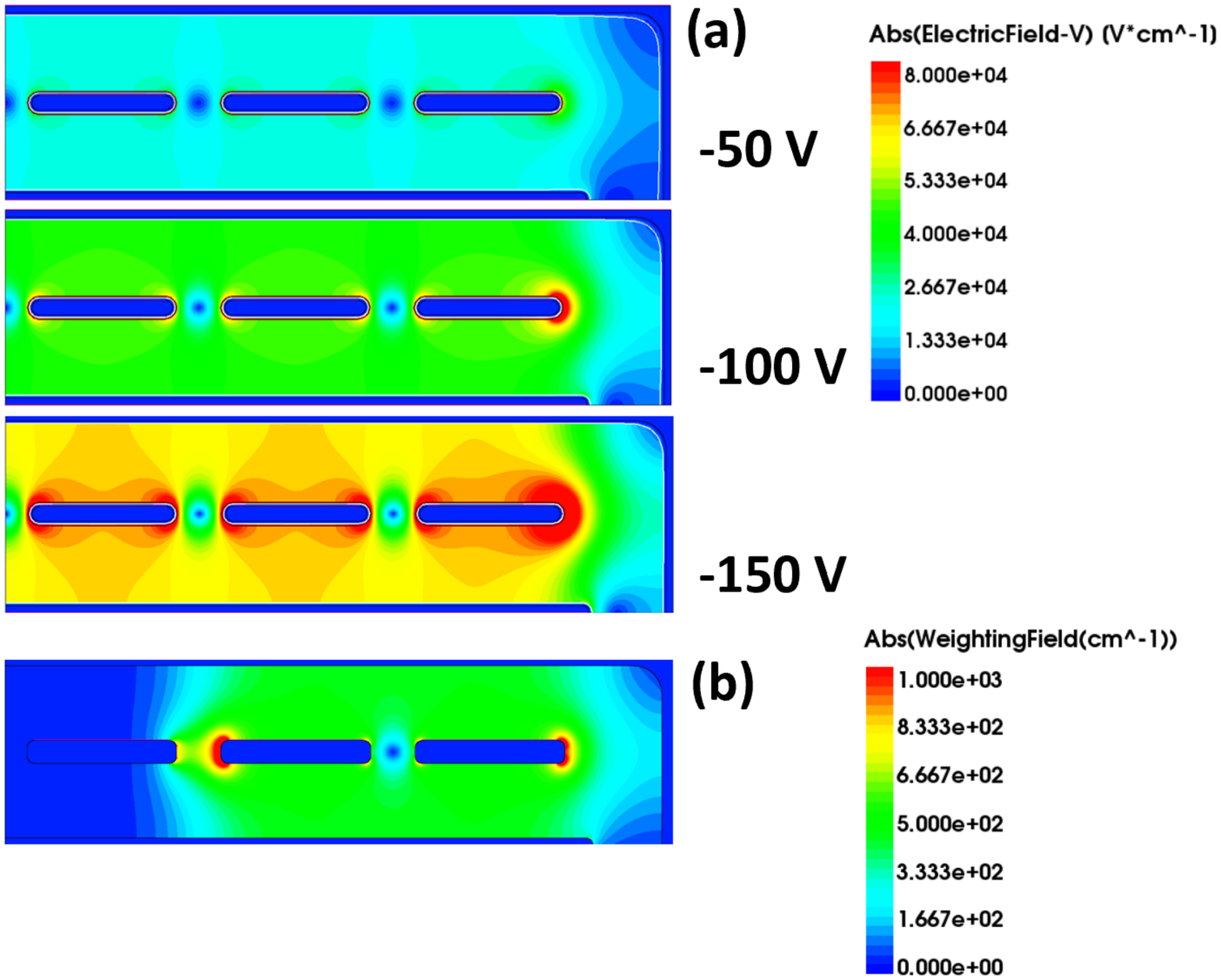} 
		\caption{(a) Electric field amplitude at different bias voltages for the double-pixel test structure and (b) weighting field.}
		\label{fig:TestStructurePhysics}
\end{figure}
The sensor design and the simulation of its physical properties (electric and weighting fields, charge carrier mobility) is performed by using Synopsys Sentaurus TCAD package~\cite{TCAD}.  
The model is simulated using a quasi-stationary voltage ramp from \SI{0}{\volt} to \SI{-150}{\volt}. The relevant information at the bias voltages of \SI{-50}{\volt}, \SI{-100}{\volt} and \SI{-150}{\volt} are saved for the subsequent transient simulation. As shown in figure~\ref{fig:TestStructurePhysics}, the simulated test structure presents an uniform electric field in the regions placed between the ohmic and readout electrodes. Areas with smaller electric field are located in the inter-pixel regions between the readout electrodes and in the active volume aside the rightmost pixel. The inter-pixel areas are not particularly critical for fast timing, thanks to the small drift path travelled by charge carriers to reach their collection electrodes and the higher weighting field, which implies stronger current induction. 
The region aside the rightmost pixel (referred to as ~\emph{side region} in the following) instead is the main critical region for the timing performances of this particular test structure. Larger distances from the electrodes and less uniform electric and weighting fields cause larger charge collection times and a more diversified current signal depending on the position.

%%%%%%%%%%%%%%%%%%%%%%%%%%%%%%%%%%%%%%%
\section{Energy deposit simulation}
The studies described in this paper rely on precise simulations of the energy deposit in silicon from both high-energy ionising particles and infrared laser beams. 
Specific simulation tools were developed to overcome the limitations of the Sentaurus TCAD package in this simulation task.

\subsection{Ionising particle simulation}\label{sec:mipdeposit}
The energy release of a minimum ionising particle (MIP) in matter is a stochastic process mainly due to ionisation, atomic excitation and production of highly ionising $\delta-$rays.

Within the Sentaurus TCAD package such processes are simulated by the \textit{HeavyIon} (HI) model, which implies several approximations. Firstly, the stochastic nature of the process is completely neglected since the total expected energy release is distributed uniformly along the particle trajectory. Secondly, this approach does not allow to describe more complex deposits, such as those due to the highly ionising $\delta-$rays. 

To overcome these limitations the GEANT4~\cite{Geant4} Monte Carlo simulator is used to model the energy deposit in the sensor.
The simulation performs a sequence of single particle interactions in the silicon detector. Each particle is a positive pion with momentum of $270$~MeV/$c$ (equal to the pion momentum at PSI test beam) 
and impinges on the detector surface with an uniform spatial distribution and with an angular distribution in agreement with the characteristics of the PSI $\pi$-M1 beam line (angular divergence on the target of \SI{35}{\milli\radian} horizontal and \SI{75}{\milli\radian} vertical).
For each event, GEANT4 saves the energy deposits and the trajectories of the incoming pion and all secondary particles produced in its interaction with the silicon detector.
This information is then used by the TCoDe simulation package (see section~\ref{sec:TCoDe}) to compute the 
charge carrier deposits and the currents induced on the readout electrode due to the drift of the charge carriers.

\subsection{Laser beam simulation}\label{subsec:laserdep}
\begin{figure}[t!]
	\centering
		\includegraphics[width=1.0\linewidth]{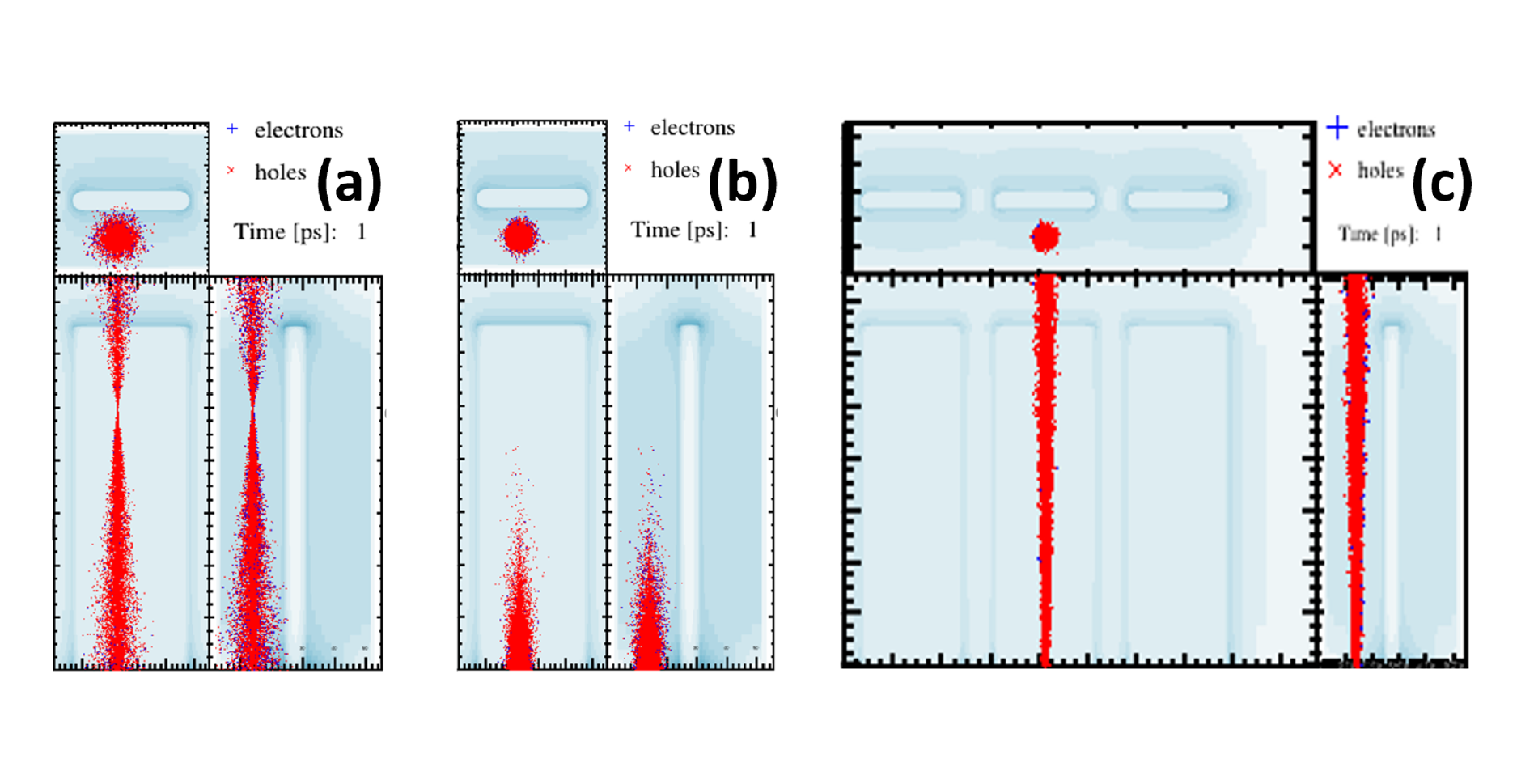}
		\vspace{-1.2cm}
		\caption{Examples of simulated energy deposit shapes from laser sources inside the TIMESPOT test structure. (a) Deposit with focus inside the active bulk. (b) Deposit shape due to high absorption (655 nm wavelength). (c) Deposit of IR laser source (1030 nm wavelength). }\label{fig:laserSimulation}
\end{figure}

In addition to the measurement at the test beam, a set of measurements  using an ultra-fast infrared (IR) laser source were performed as detailed in section~\ref{subsec:semiempiricalmethod}. The energy deposition from laser pulses has been simulated with a specific tool developed and implemented within the TCoDe package. 
The algorithm takes into account the Gaussian shape of the laser beam, the used wavelength, the light absorption in the material and the position of the focus. 
For each photon of the laser beam, electron-hole pairs are generated along the laser direction at a depth that follows an exponential distribution, to account for the light absorption. 
Figure~\ref{fig:laserSimulation} shows the different projections of the charge carrier distributions generated by different laser sources. 

%%%%%%%%%%%%%%%%%%%%%%%%%%%%%%%%%%%%%
\section{Dynamic modelling and transport mechanisms: TCoDe}\label{sec:TCoDe}
Albeit usable in principle, the TCAD software turned out to be inadequate to perform an exhaustive 3D full-volume simulation of the induced current signals at the electrodes starting from the simulated charge deposits. The complete simulation of one induced current signal from a single track in a 55$\times$55$\times 150 \,\mu$m$^3$ volume takes at least 30 hours on a 24-core machine. 
Moreover, TCAD simulation results are extremely sensitive to the applied mesh strategy, which has to be adapted to the corresponding drift path of the charges in order to avoid numerical errors such as numerical diffusion. 
For these reasons a dedicated simulation algorithm has been developed (TIMESPOT  Code for Detector simulation, TCoDe~\cite{TCode}) to calculate separately the effect of single charge trajectories inside the volume and the consequent induced signal. This approach is completely different with respect to the one used by TCAD, which is based on the collective movements of the charge carriers within a charge cloud. The TCoDe procedure allows a very efficient parallelisation in the calculation of the charge carrier dynamics with a huge gain in terms of simulation speed, using an algorithm that can run both on CPU (single-thread) and GPU (multi-thread). As an example, a simulation run of 50 different tracks crossing the volume of three adjacent pixels would take three months using TCAD, while TCoDe completes it in 1 hour and 30 minutes in single-thread mode (on a Intel Xeon CPU X5450, 10 GB RAM) and in 1 minute and 40 seconds on a commercial gaming laptop on GPU in multi-thread mode. This feature allows to simulate large event samples needed for detailed studies of the sensor performances.

The TCoDe simulator receives in input the TCAD-generated physics maps, containing spatial information of the charge carrier mobility, the electric and weighting field of the interested device, and the energy deposits in the sensor material obtained from GEANT4 or the laser simulation. The output consists of the induced currents signals for the electron and hole contributions as well as for primary and secondary particles. 
Samples of $50\,000$ events are generated for each simulated configuration. 

\section{Front-end response simulation: TFBoost}
The sensor readout electronics used in the 2019 PSI test beam is based on a two-stage signal amplification scheme acting as an inverting transimpedance amplifier, implemented on a custom-made circuit and extensively described in ref.~\cite{JINST-TimeSpot}.

For a quantitative comparison with measurements, the induced signals simulated with TCoDe need to be convoluted with such front-end electronics response. In order to accomplish this, an application named TFBoost (TIMESPOT Front-End Booster)~\cite{TFBoost_cit} has been developed, based on the HYDRA framework~\cite{hydra}, implementing complete modellisation of the front-end electronics response and its corresponding noise. In particular, TFBoost allows to perform convolutions between the set of input signals from TCoDe and the electronics transfer function in an extremely efficient way by exploiting multi-thread parallelism, both in CPU and GPU. As an example, in TFBoost a single convolution between two signals with 1 million sample points, computed in frequency domain, would take approximately 1 second in single thread-mode and 330 ms in multi-thread mode, on a Intel Xeon Silver CPU 4116, while it takes 130 ms using an NVIDIA A100 PCIe GPU. The same computation made in time domain, with a non optimised single-thread algorithm, would take approximately 1 hour (on Intel Xeon CPU 4116). Thus, like in TCoDe, TFBoost allows to process and analyse large samples of induced current signals.

\subsection{Simulation flow}
\label{subsec:tfboost_flow}
TFBoost parses the output signals from TCoDe and computes the convolution with a specific front-end transfer function, either provided by the user or chosen within the TFBoost library. The signals are defined in a maximum time range of 100~ns, with an original time step of 1~ps, resulting in $100\,000$ sample points. As a result of the multi-thread parallelism, convolution computation for a set of $50\,000$ signals takes few minutes.
The output signals are then re-sampled with a time step of 20~ps, emulating the time digitisation of the oscilloscope used during the test beam. The noise is then added to the signal in two possible ways: noise samples can be computed from a configurable noise analytical model, or they can be provided externally by the user, for example when noise samples are available as a set of experimental measurements. A detailed description of the noise and transfer function used in this analysis is given in the following sections.
Finally, the voltage values of the output signal are digitised simulating the 8 bit ADC of the oscilloscope used in the test beam. The ADC digitisation is applied to the noisy signal if noise is simulated from the analytical model, or to the noise-less signal if the measured noise is used.
Only simulated signals with an amplitude larger than a threshold value are saved for the analysis. This value reproduces the data acquisition threshold used at the test beam.

\begin{figure}[thb!]
    \centering
    \includegraphics[scale=0.75]{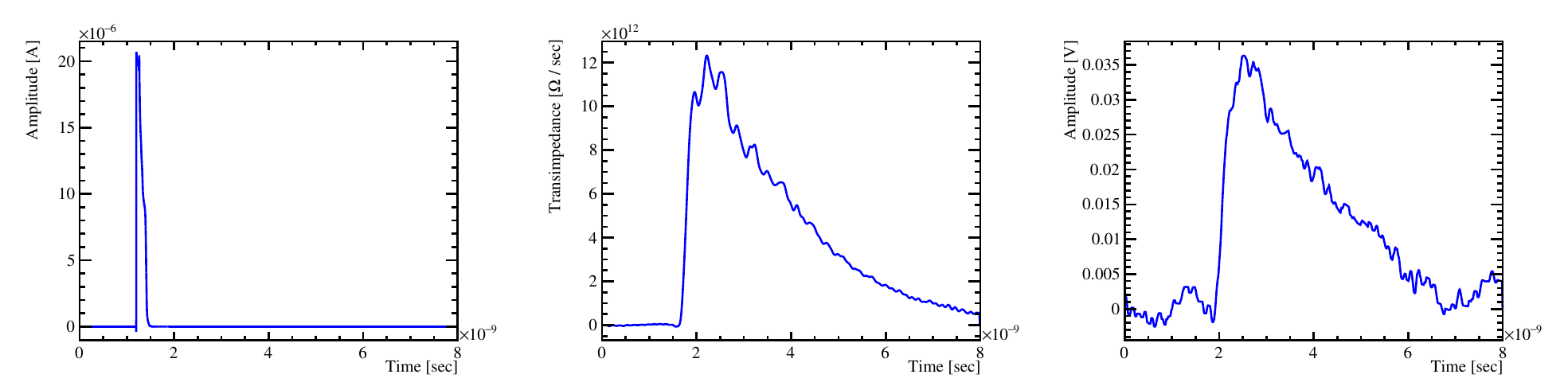}
    \caption{Example of the result of the front-end simulation for a single input current from the double pixel test structure, at $-150$~V bias voltage. (Left) input current for a MIP deposition in the sensor, (centre) simulated transimpedance and (right) output signal waveform with real noise. 
    }
    \label{fig:conv}
\end{figure}

Each simulation step can be individually turned off in order to study in detail the sensor and front-end electronics performances.
An example of the result of the front-end simulation for a single input current is shown in figure~\ref{fig:conv}.

\subsection{Semiempirical method for transfer function determination}
\label{subsec:semiempiricalmethod}
\begin{figure}[b!]
    \centering
    \includegraphics[scale=0.75]{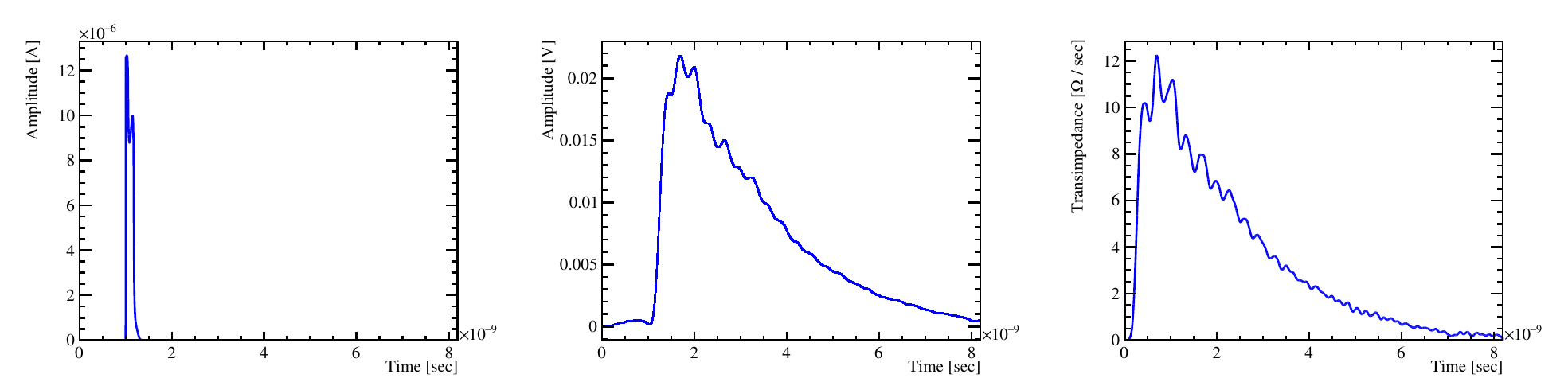}
    \caption{Example of the result of the front-end transfer function deconvolution with the semiempirical approach. (Left) Input current for an IR laser deposition in the sensor. (Centre) Averaged measured waveform using the IR laser setup. (Right) Deconvoluted front-end transfer function.}
    \label{fig:deconv}
\end{figure}

The transfer function acting on the transient signals generated in the sensor is characterised not only by the front-end electronics itself but also by the sensor capacitance and the impedance of the sensor-electronics connection (e.g. wire bonding). For this reason, a direct and accurate measurement of such transfer function requires the sensor connected to the electronics.
Consequently the main method for simulating the transfer function consists in a semiempirical approach. The double-pixel TIMESPOT sensor is irradiated using a 200~fs width, 1030~nm wavelength laser with a minimum spot size on target of $5\, \mu$m~\cite{IRLASER}. The laser intensity is adjusted to obtain an energy deposition in the sensor corresponding to 1~MIP (approximately 2~fC), and the sensor is read out with the same front-end electronics used at the test beam. The output signals are obtained by averaging 3000 single waveforms to suppress the noise and reach a signal-to-noise ratio (SNR) of about 54~dB. The laser irradiation is repeated in different positions within the active area of the sensor and at different bias voltages. 
The output signals of the front-end electronics are then deconvoluted with TFBoost using the simulated currents obtained from TCoDe at the corresponding laser positions and bias voltages, as described in section~\ref{subsec:laserdep}.
If the current transients of the sensor are precisely simulated, the deconvoluted responses are semiempirical precise descriptions of the real front-end transfer function~\cite{SCHARF20151}.
An example of the deconvolution procedure between one input TCoDe current and one measured averaged waveform is shown in figure~\ref{fig:deconv} at $-150$~V bias voltage and in one specific irradiation position.

Several sanity checks are performed in order to assure the absence of instabilities in the semiempirical deconvolution procedure. A white noise is added to the output signal before performing the deconvolution with different RMS levels: no difference in the resulting transfer function shapes are observed until reaching SNR of 29~dB, compatible with similar results that can be found in ref.~\cite{deconv_noise}. Moreover, time and frequency aliasing effect in the deconvoluted transfer function are reduced to a negligible level by using a time window greater than 10~ns and by interpolating the output voltage signal with a cubic spline and by resampling it with a 1~ps time step. The presence of well-known spurious high-frequency spikes in the deconvolution result~\cite{deconv_problem} are suppressed to a negligible level using a post-deconvolution low-pass Butterworth filter of 10th order, with cutoff frequency of 10~GHz, following methods described in refs.~\cite{deconv_noise,deconv_problem}.

\begin{figure}[t!]
    %\centering
    \begin{minipage}{0.46\textwidth}
    \includegraphics[scale=0.34]{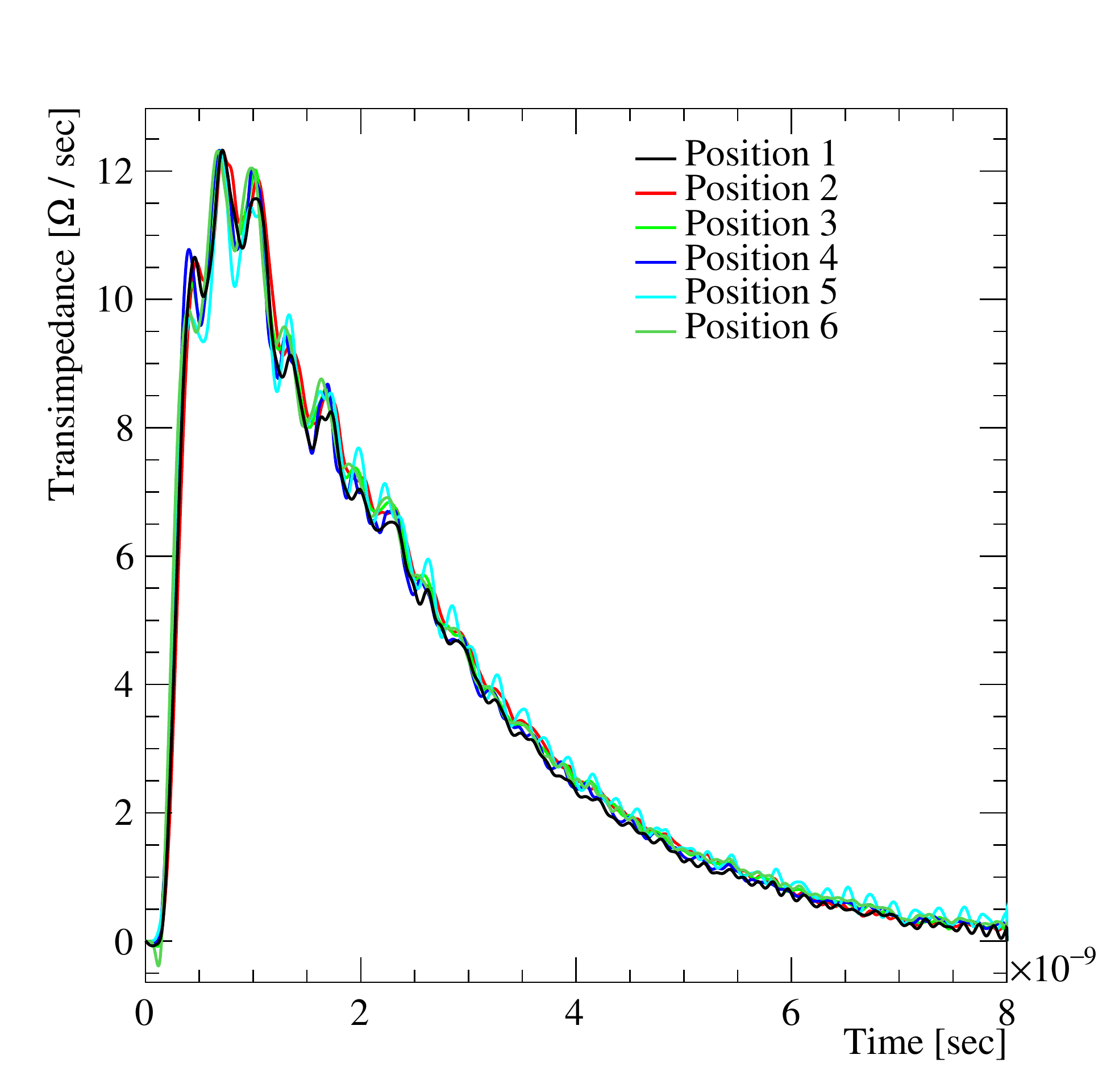}
    \end{minipage}
    \begin{minipage}{0.46\textwidth}
    \includegraphics[scale=0.275]{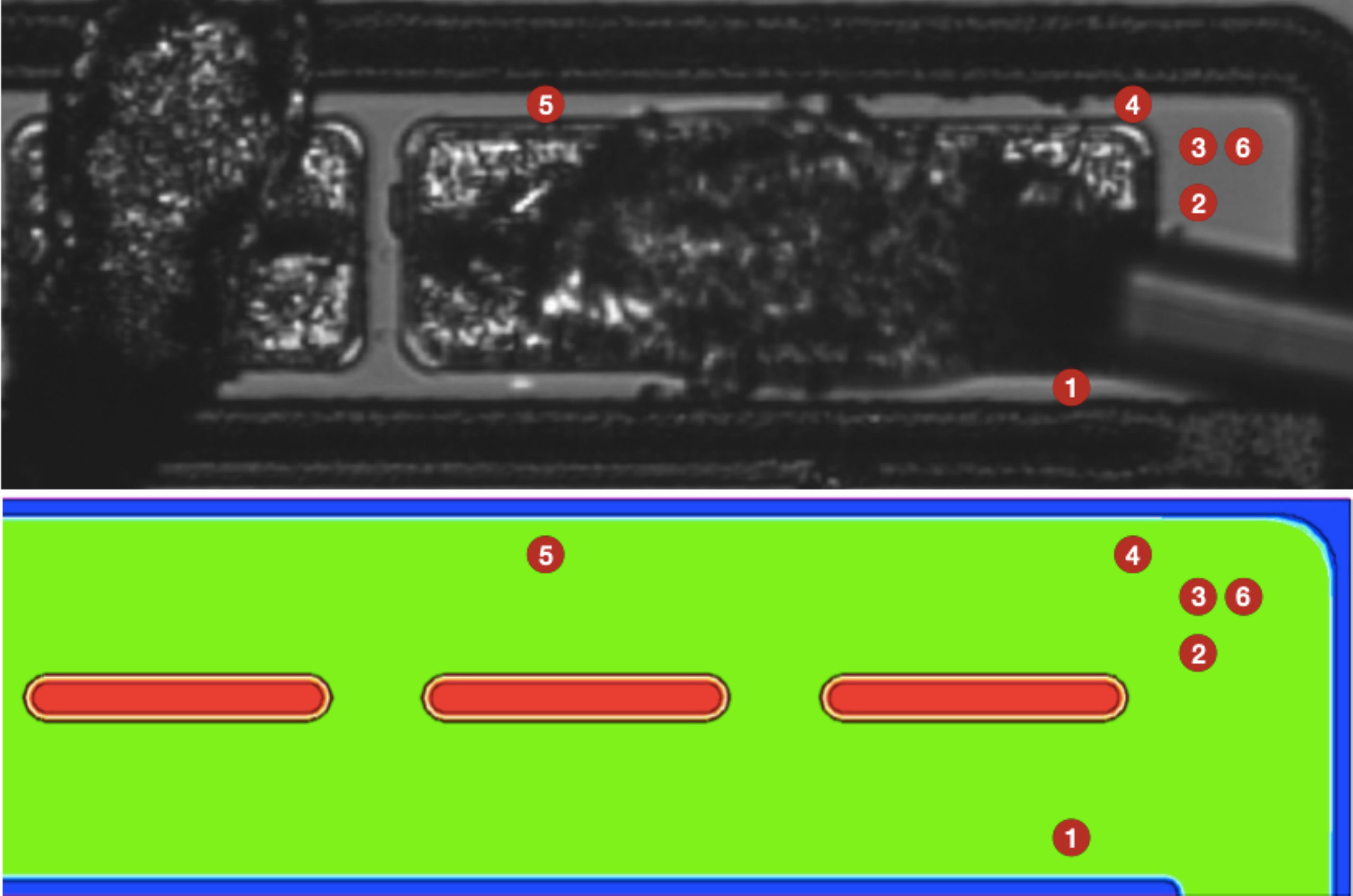}
    \end{minipage}
    %\vspace{-0.5cm}
    \caption{(Left) Comparison of semiempirical transfer functions obtained in different irradiation positions with the laser setup, for the $-150$~V bias sample. (Right) Illustration of the six irradiation positions within the active area of the actual double-pixel test structure (top) and the corresponding positions in the simulated structure (bottom).}
    \label{fig:multipletflasers}
\end{figure}

In order to verify the level of approximation in the description of the real electronics response, a consistency check is performed: since ideally the semiempirical transfer function must depend only on the electronics response, the sensor electrical properties and the electronics-sensor coupling, a direct comparison of the functions obtained at different laser irradiation positions and at different bias voltages is done, resulting in a very good agreement, especially in the rising edge. An example of such comparison for the $-150$~V bias sample is shown in figure~\ref{fig:multipletflasers}. The transfer function obtained in position 1 is chosen as the final one to be used in the subsequent analysis, because it corresponds to the fastest transient current in the sensor for the available irradiated positions, leading to the most accurate determination of the frequency components of the deconvoluted front-end response.

\subsection{Analytical transfer function}
\begin{figure}[b!]
    \centering
    \includegraphics[scale=0.34]{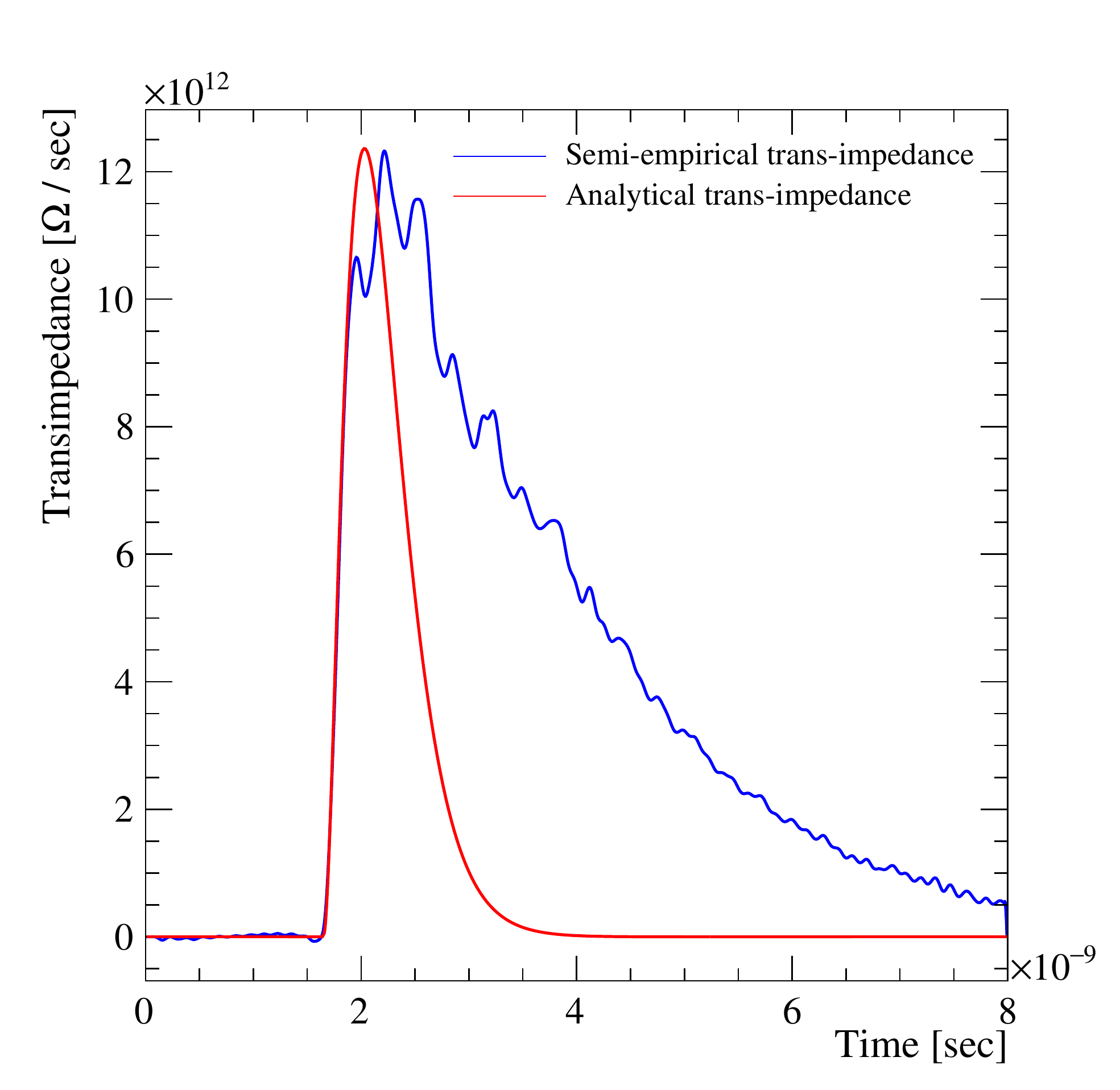}
    \vspace{-0.5cm}
    \caption{Comparison between the analytical and semiempirical transfer functions.}
    \label{fig:analyticaltr}
\end{figure}

As a consistency check, a second method for simulating the front-end electronics is considered, using a fully analytical model. 
It consists of modelling the first amplification stage as a second order transimpedance transfer function with DC transimpedance $R_{m_0}$, as described in ref.~\cite{FastTiming}, while the second amplification stage is described as a single pole, inverting, voltage gain transfer function with DC voltage gain $G_0$. The inverse Laplace transform of the overall transfer function reads
\begin{equation}
\mathcal{R}(t)=\mathcal{L}^{-1}(t)  \Bigg\{-\frac{R_{m_0}}{(1+s\tau)^2} \frac{G_0}{1+s\tau^{*}} \Bigg\},
\end{equation}
\begin{equation}
\mathcal{R}(t)=-G_0R_{m_{0}} \Bigg\{ \frac{(t(\tau-\tau^{*})-\tau \tau^{*})}{\tau(\tau-\tau^{*})^2}e^{-\frac{t}{\tau}} - \frac{\tau^{*}}{(\tau-\tau^{*})^2}e^{-\frac{t}{\tau^{*}}}\Bigg\},
\end{equation}\\
where $\tau$ and $\tau^{*}$ are the time constants of the first and second stage respectively. The first stage is based on a board developed at the University of California Santa Cruz.  
Because of a different high-frequency feedback path with respect to the one used in the model and the presence of a load inductance, a more precise analytical description would require at least five poles. The proposed model still allows to obtain a reasonable description of the behaviour of the electronics, once the time constants and the DC transimpedance $ G_0R_{m_{0}}$ are accurately calibrated. 
In order to accomplish this, the analytical transimpedance (defined in the TFBoost library) is convoluted with the TCoDe currents in the case of laser irradiation in different positions of the sensor. The corresponding output signals are then compared with the averaged waveforms measured using the laser setup, and the transimpedance parameters are tuned in order to match the voltage output amplitude and rise time.  
Figure ~\ref{fig:analyticaltr} shows a comparison of the semiempirical and analytical transimpedance, which have a very similar rising edge. Therefore similar timing results are expected when using the analytical transimpedance in the simulation. Nevertheless, the advantage of the semiempirical method consists in describing the entire signal waveform, thus leading to a better agreement between simulation and test beam results. Indeed, the simulation data analysis and the comparison with test beam experimental results have been repeated using the analytical electronics response, obtaining consistent results to those discussed in section~\ref{sec:results}.

\subsection{Noise}
\label{subsec:noise}
Similarly to the electronics response study, two methods are followed for the noise modelling: use of noise samples measured at test beam and generation of red noise samples using an analytical model.

The true noise-only waveforms were acquired during the test beam with a sampling time step of 20 ps and at different bias voltages. Each waveform is added to each simulated output signal, after the time digitisation step in TFBoost, as described in section~\ref{subsec:tfboost_flow}, considering the corresponding bias.
This procedure allows to take into account possible extrinsic noise sources, which are unknown and thus impossible to be properly described in the simulation, and is therefore chosen to be the main procedure followed in this analysis. Nevertheless, in order to get a deeper comprehension of the noise characteristics and not to be limited by the finite statistics of the measured noise waveforms, the definition of an analytical model for noise generation is also investigated.
 
\begin{figure}[thb!]
    \centering
    \includegraphics[width=0.45\textwidth]{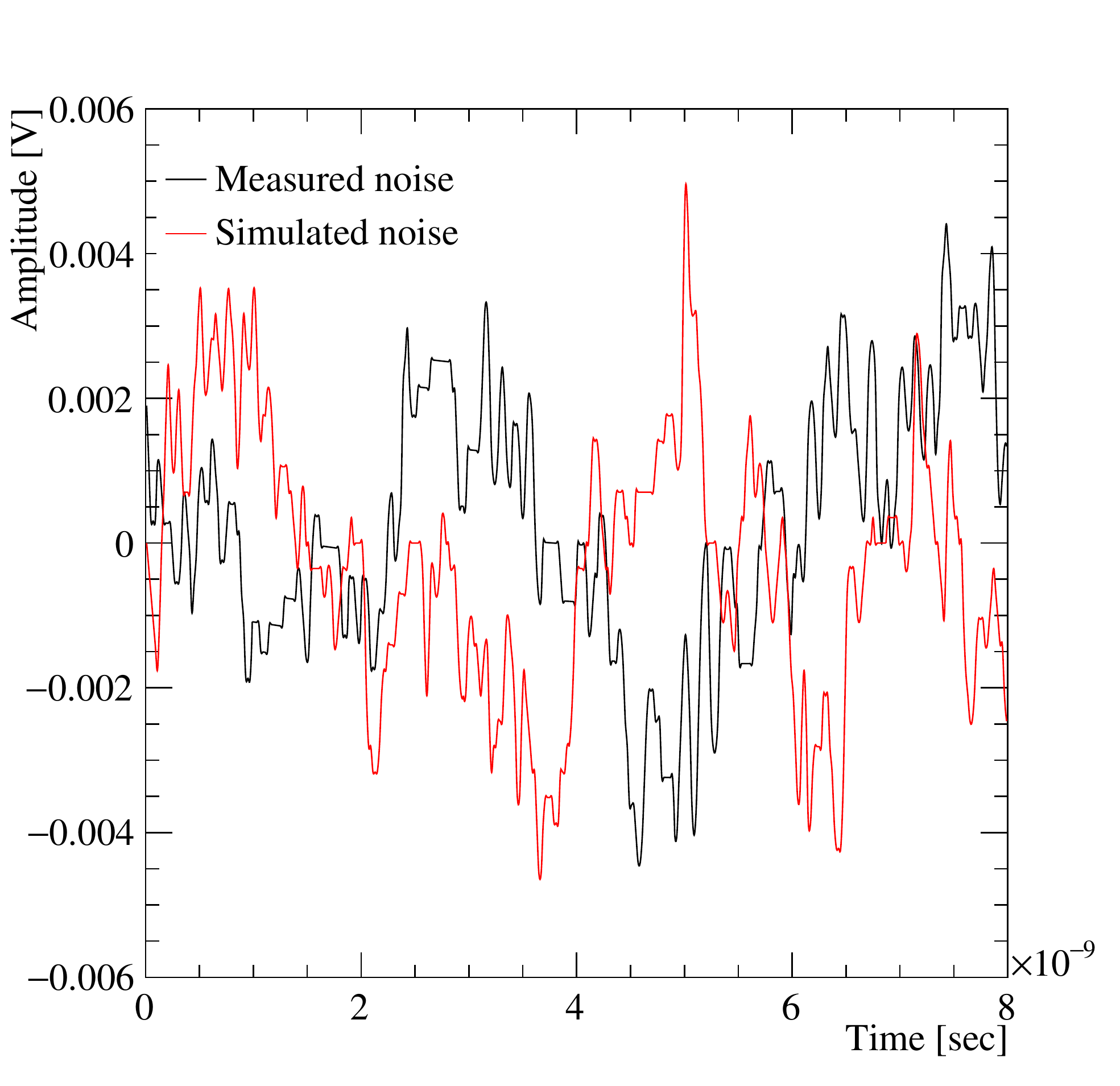}
    \includegraphics[width=0.45\textwidth]{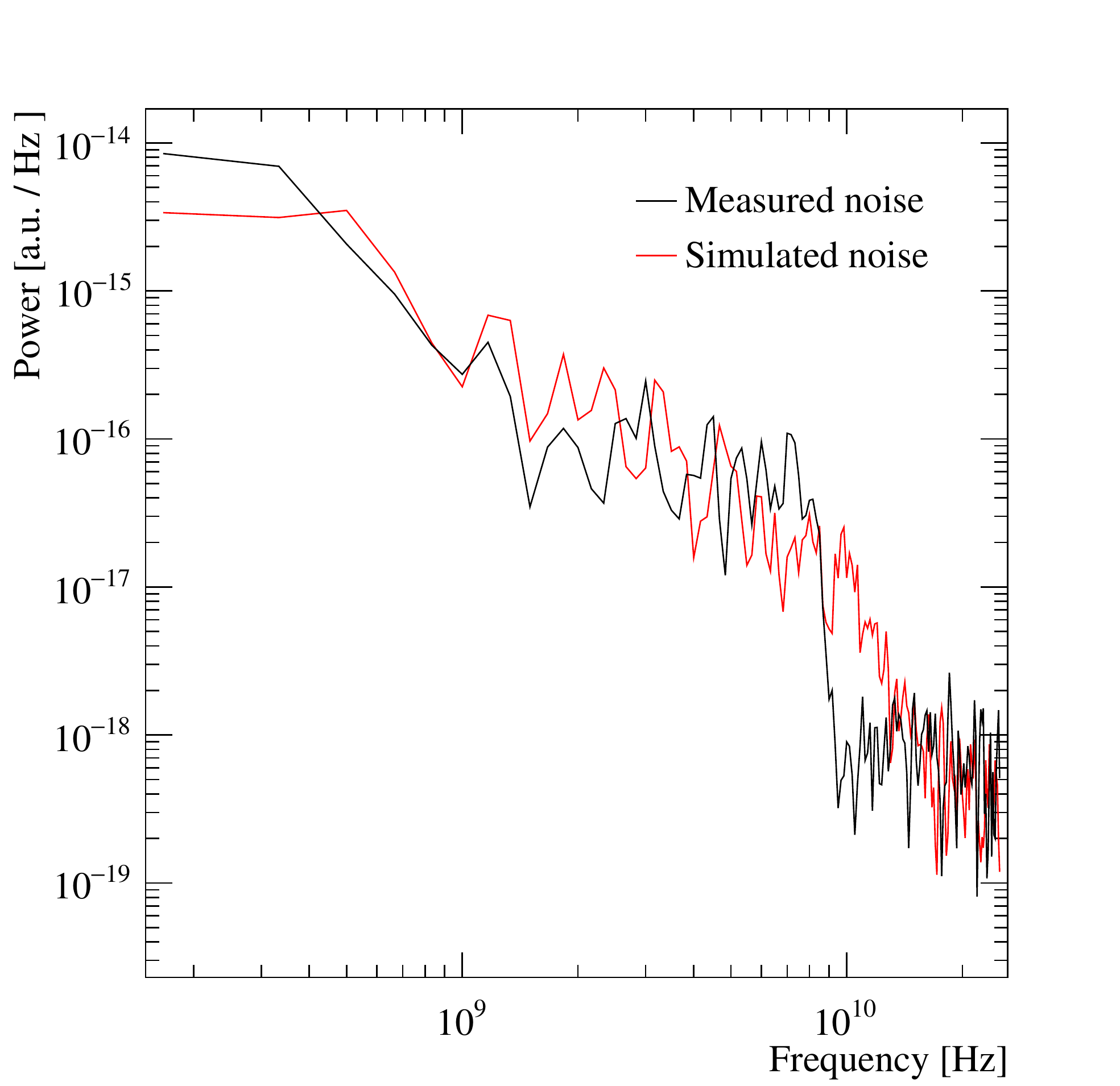}
    \vspace{-0.2cm}
    \caption{Comparison between two waveforms of  (black) measured and (red) simulated noise for the $-150$~V bias sample. (Left) Noise waveform in time domain, (right) power spectral densities.}
    \label{fig:bothnoise}
\end{figure}

By analysing the power spectral density of the measured noise, shown in figure~\ref{fig:bothnoise} for the $-150$~V bias sample, it must be noted that it is not uniformly spread across all frequencies, but shows higher intensity at lower frequencies. Therefore the noise can be described with a red noise analytical model. Specifically, red noise samples are generated from an original white noise sequence following ref.~\cite{red_noise_lect}, and introducing a correlation $r$ between samples, with $0<r<1$, where $-\log(r)/\delta T$ is the decorrelation rate and $\delta T$ is the time step between the samples. Red noise points are then calculated as 
\begin{equation}
    x_i = r\,x_{i-1} + \sqrt{1-r^2}\,y_i,
\end{equation}
where $x_i$ and $y_i$ are the i-th red and white noise samples respectively. The coefficient $r$ is tuned in order to reproduce the main trend of the experimental noise spectrum. In particular for the front-end electronics used at test beam for the $-150$~V bias sample the correlation is $r=0.9851$, which corresponds to a $0.015 \,\text{ps}^{-1}$ decorrelation rate, or equivalently to a decorrelation time of 67~ps. It can be demonstrated that the original white noise and the resulting red noise have the same RMS, which has been tuned to $2.2 \,\text{mV}$. Similar results are obtained for the other bias voltage samples.
Figure~\ref{fig:bothnoise} shows a comparison between the measured and simulated noise for the $-150$~V bias sample, in time and frequency domain. The analysis and the comparison with test beam results have been repeated using the simulated red noise, obtaining equivalent results to those discussed in section~\ref{sec:results}.

\section{Results}\label{sec:results}

\subsection{Comparison with test beam measurements}
As discussed in section~\ref{sec:intro}, the 3D-trench pixel geometry leads to uniform signal shapes and narrow distribution of the charge collection times at the electrodes.
Minor nonuniformities of the electric field exist and are due to the limited size of the readout trench. 
Moreover, in the test structure studied in this paper, about 20\% of area next to the double pixel is active and characterised by a smaller, but nonzero electric field.
Finally, the difference in the charge carriers drift velocities introduces some nonuniformities in the signal shapes depending on the position where signal originates. All these aspects determine the properties of the specific silicon detector studied in this paper.
In order to compare these properties, resulting from simulation, with the results from the 2019 test beam~\cite{JINST-TimeSpot}, simulated waveforms of the 3D-trench test structure are analysed using the same procedure applied to data detailed in ref.~\cite{JINST-TimeSpot}.

\subsubsection{Waveform properties and amplitude}

\begin{figure}[b!]
    \centering
    \includegraphics[width=0.45\textwidth]{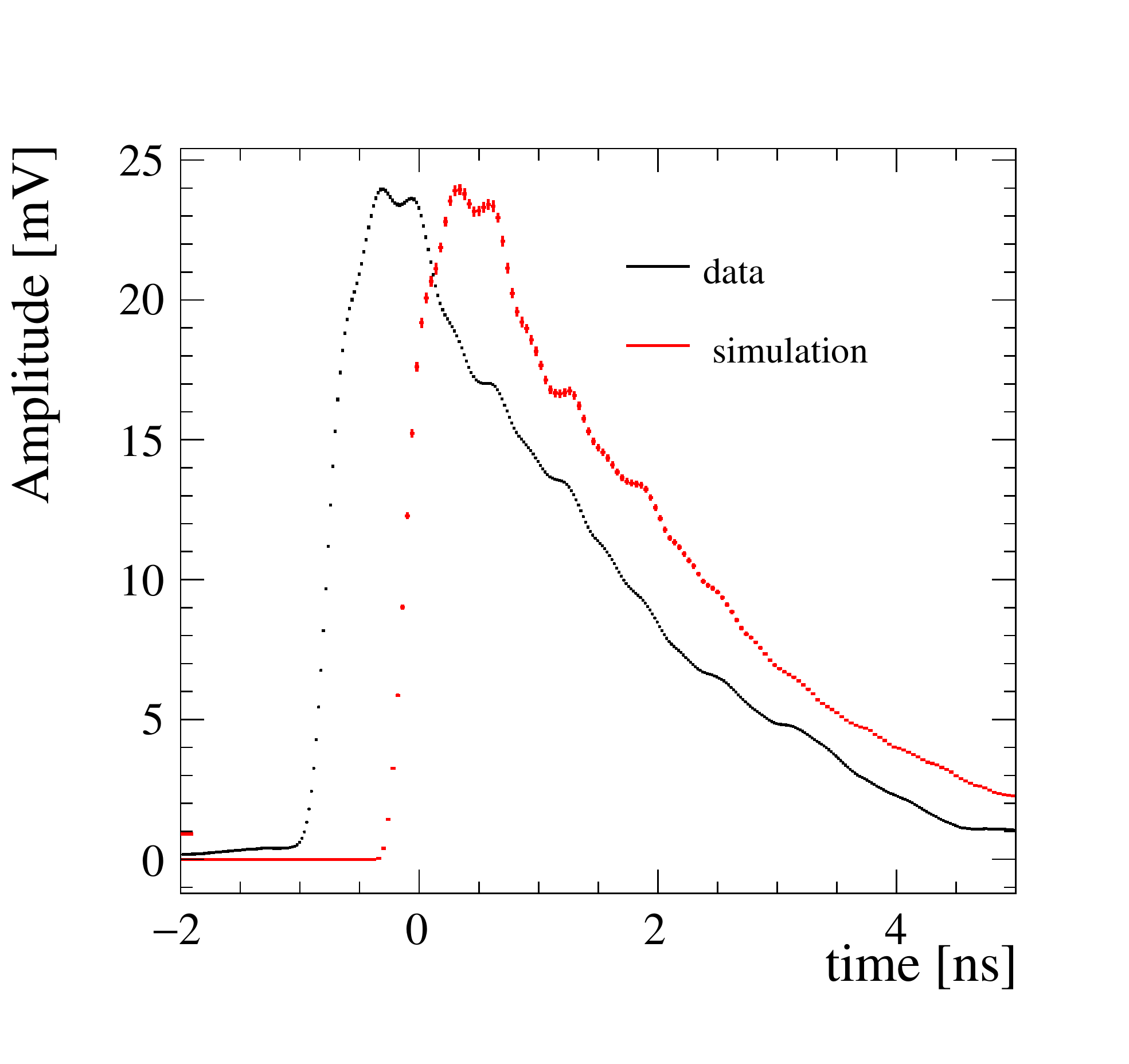}
    \caption{Silicon sensor average waveform from (black) data and (red) simulation. An arbitrary time shift between the two shapes is applied to allow a qualitative comparison.}
    \label{fig:Profile}
\end{figure}

\begin{figure}[thb!]
    \centering
    \includegraphics[width=0.45\textwidth]{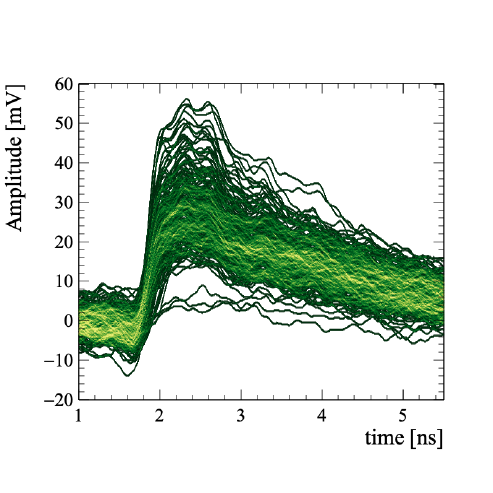}
    \includegraphics[width=0.45\textwidth]{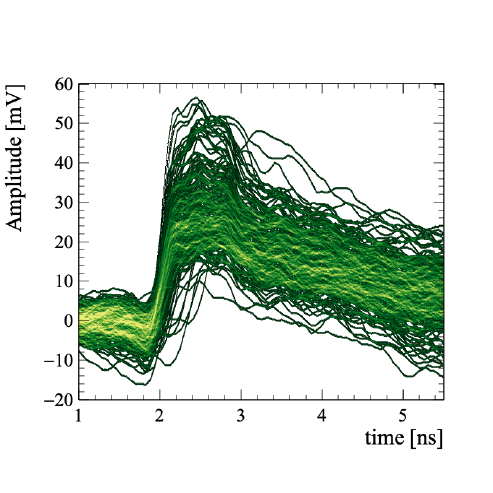}
    \caption{Overlap of 200 silicon sensor waveforms for (left) simulation and (right) test beam data.}
    \label{fig:Profile2}
\end{figure}

%%%%%%%%%%%%%%%%%%%%%%%%
%% GENERAL PARAMETERS %%
%%%%%%%%%%%%%%%%%%%%%%%%
The use of the semiempirical transimpedance, described in section~\ref{subsec:semiempiricalmethod}, allows to reproduce with very good accuracy the different structures visible in the average waveform shown in figure~\ref{fig:Profile}. The good agreement between data and simulation is also visible by comparing qualitatively the single waveforms, as shown in figure~\ref{fig:Profile2}, in terms of amplitude, rise time and noise fluctuations. The main quantities representing the signal properties agree within 5\% and are summarised in table~\ref{tab:properties}.

\begin{table}[h]
\centering
\caption{Maximum amplitude, average signal-to-noise ratio, noise, rise time (20-80\%) and slew rate (dV/dt) of the 3D-trench silicon sensor response at different values of the bias for simulation and data. The statistical uncertainties are below 1\%.}\label{tab:properties}
\vspace{0.3cm}
\resizebox{0.7\columnwidth}{!}{
\begin{tabular}{| l | c  | c  c c c c |}
\hline
& $V_{\rm bias}$   & Amp(P$_{\rm max}$)  & $\langle$S/N$\rangle$ & $\langle$N$\rangle$ & rise time & dV/dt \\
& [V]   &  [mV] &   &  [mV] &   [ps] &   [mV/ns]\\ \hline
\multirow{3}{*}{Simulation}
& $-50$ & 25.0 &  14.6 & 2.11 & 247 & 103  \\
& $-100$ & 24.5 & 14.3 & 2.17 & 224 & 113 \\
& $-150$ & 24.4 & 14.2 & 2.19 & 217 & 116 \\
\hline
\multirow{3}{*}{Data }
& $-50$ 	  	 & 24.1  & 14.3 & 2.19 & 258 &  111 \\
& $-110$ 	  	 & 24.4  & 13.9 & 2.30 & 221 &  123 \\
& $-140$ 	  	 & 24.7  & 14.2 & 2.29 & 217 &  126 \\
\hline
\end{tabular}
}
\end{table}
\begin{figure}[h!]
    \centering
    \includegraphics[width=0.45\textwidth]{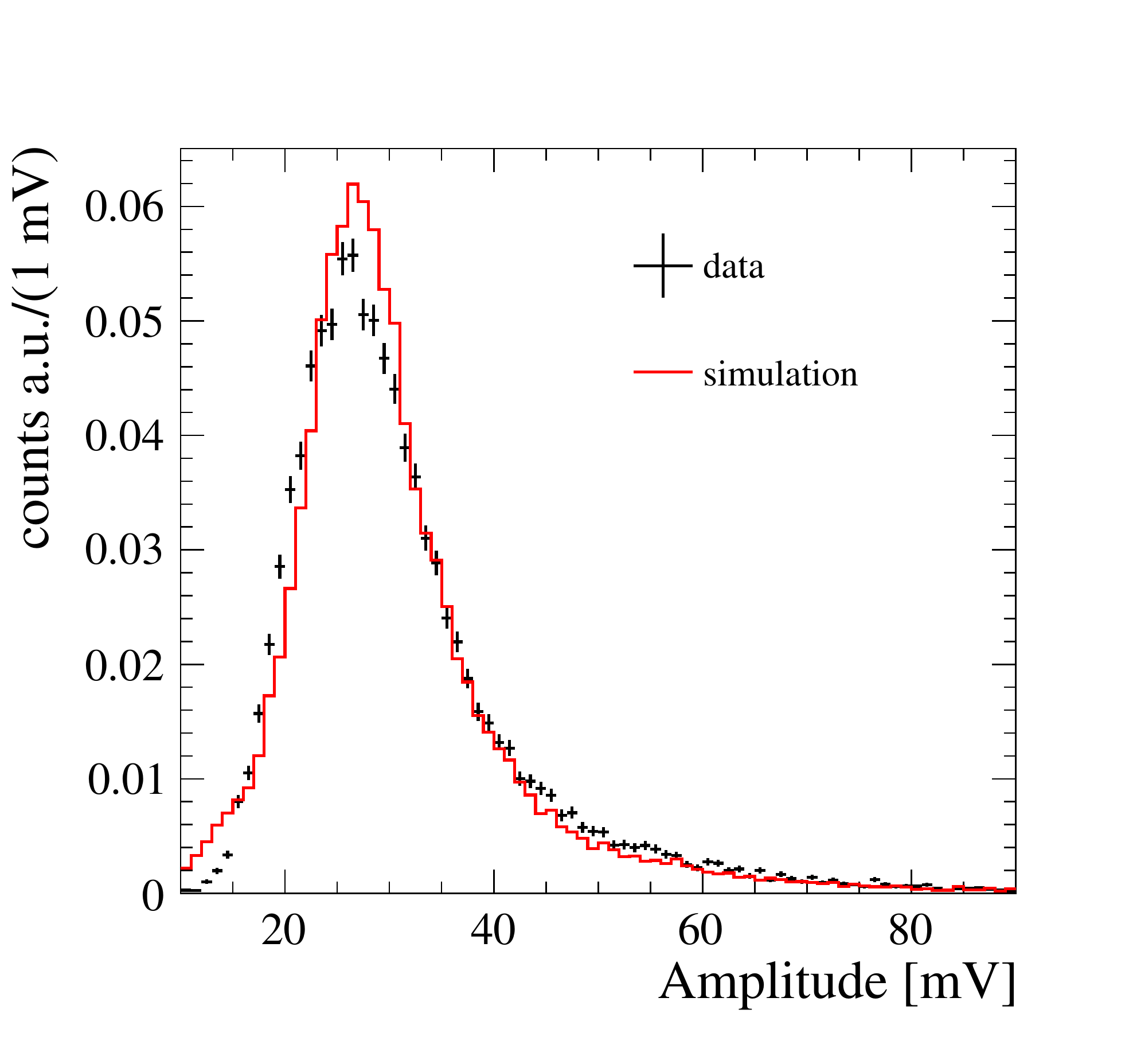}
    \caption{Distribution of the reconstructed amplitudes for the silicon sensor in data and simulation at $V_{\rm bias} = -150$~V.} 
    \label{fig:Landau}
\end{figure}

Figure~\ref{fig:Landau} shows the reconstructed amplitude for data (black) and simulation (red) at $V_{\rm bias} = -150$~V. 
The simulation reproduces the data distribution, characterised by a Landau probability density function shaped by the trigger acceptance function at low amplitudes. Residual differences between data and simulation are present and are proved not to affect the results that are discussed in the following.

\subsubsection{Time of Arrival}\label{sub:TOA}

The time measurements performed at PSI on the sensor are described in detail in ref.~\cite{JINST-TimeSpot} and updated in this paper after a revised analysis. 
In absence of an external time reference of adequate precision, 
the pion arrival time is given by the average time of two MCP-PMTs signals.
Its accuracy, of about 12.5~ps, is estimated from the width of the distribution of the time difference between the two MCP-PMTs, considering similar resolutions of the two and assuming no correlations among the signals.

The time of arrival (TOA) of the silicon sensor and its corresponding resolution are determined by means of amplitude and rise time-compensated (ARC) method~\cite{ARCmethod}. According to this method (referred to as \emph{reference} method in ref.~\cite{JINST-TimeSpot}) 
the signal waveform is processed by subtracting to it an identical contribution delayed  by about half of the signal's rise time. 
The resulting waveform, showing a peaking structure, is fitted with a Gaussian function to determine the amplitude, and the value corresponding to the 50\% of the Gaussian's amplitude is defined as the signal TOA. 

\begin{figure}[t!]
    \centering
    \includegraphics[width=0.45\textwidth]{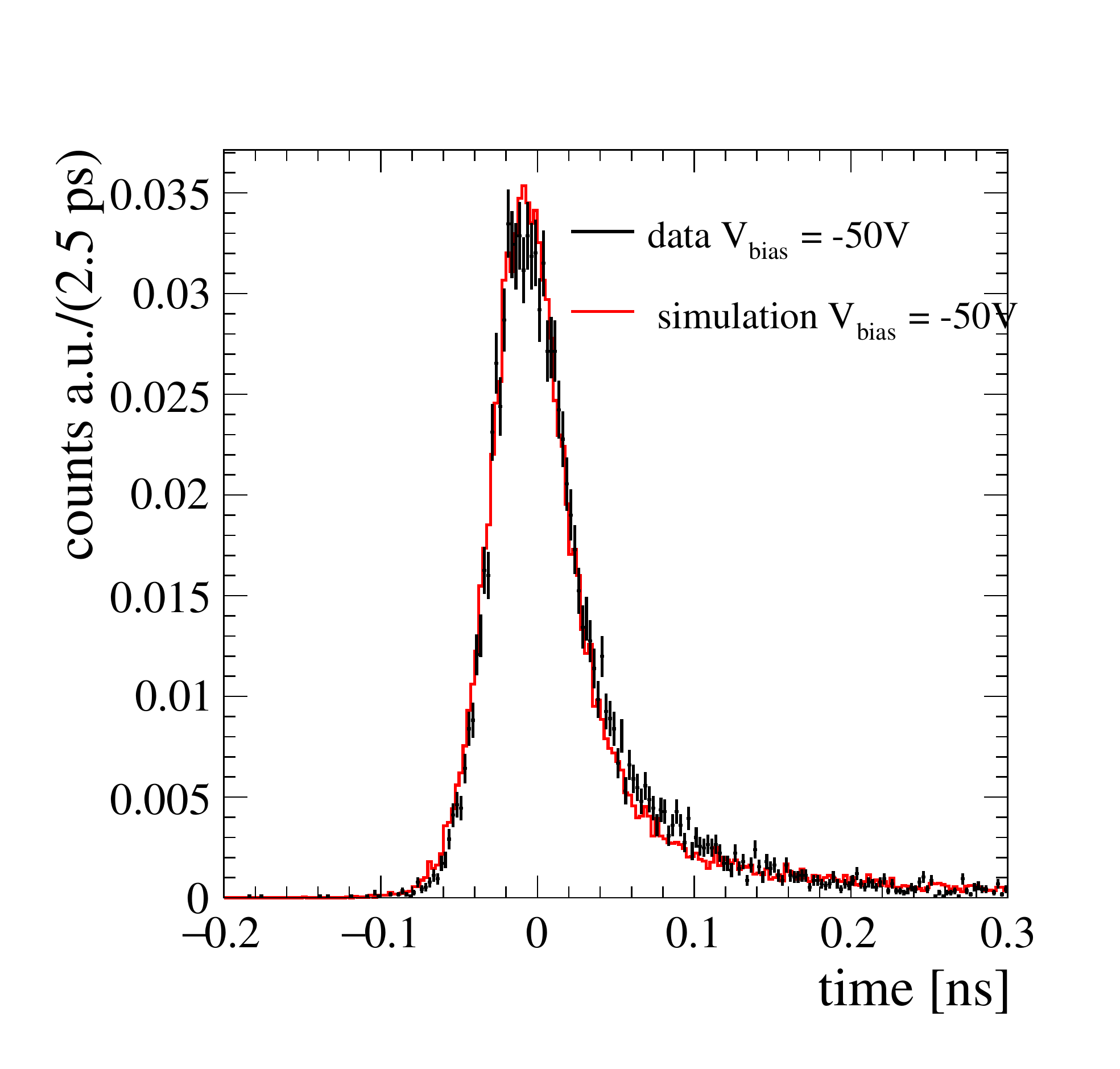}
    \includegraphics[width=0.45\textwidth]{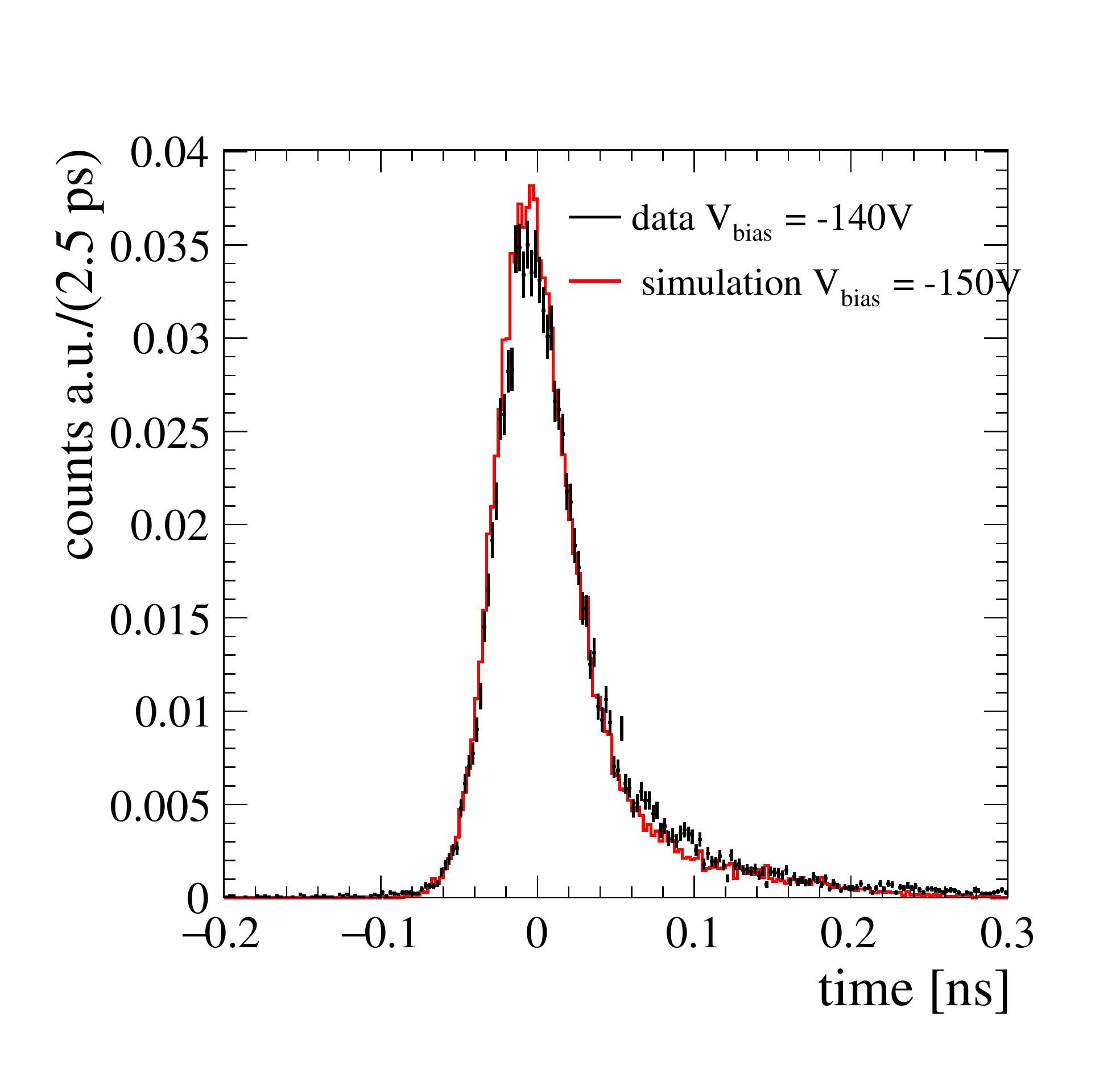}
    \caption{The distribution of the time difference between the 3D-trench silicon sensor signal and the pion arrival time at (left) $V_{\rm bias} = -50$~V and (right) $-150$~V for (black) data and (red) simulation. In the simulation the uncertainty in the time reference is accounted by adding to the time of arrival of the silicon a random value generated according to the measured resolution.}
    \label{fig:SummaryResults_timeResolution-data}
\end{figure}

Figure~\ref{fig:SummaryResults_timeResolution-data} shows the distribution of the time difference between the silicon sensor signal and the pion arrival time, $t_{\rm Si}-\langle t_{\rm MCP-PMT}\rangle$, for data (black) and simulation (red). In the simulation the uncertainty in the time reference is accounted by adding to the TOA of the silicon sensor a random value generated according to the measured time reference resolution (12.5~ps).
The distributions have a dominant peaking structure and an exponential tail of late signals. %It has been verified in simulation that the tail is entirely due to signals originated in the side region, while the signals originated in the double pixel contribute to the peaking structure.
The two distributions are in very good agreement, both in the region of the peak and of the tail. A detailed study of the tail of late signals is reported in section~\ref{sec:TimeResolutionSimulationDistr}.

A summary of the time resolution values of the silicon sensor obtained from simulation and from test beam data at different bias voltages are reported in table~\ref{tab:TimeResolution}, where the time reference resolution is subtracted in quadrature assuming no correlations among the acquired signals. Two parameters representing the time resolution are quoted for reference: $\sigma_{\rm core}$ and $\Delta_{68\%}$. The sigma of the Gaussian core of the distribution $\sigma_{\rm core}$ is obtained with a fit to the full time distribution using the sum of a Gaussian and an exponentially-modified Gaussian as probability density function. The estimator $\Delta_{68\%}$, instead, is introduced to take into account the asymmetry of the distribution in the resolution of the double pixel. The latter is obtained by taking half the smallest interval around the peaking value of the distribution that contains 68\%  of the statistics of the events originated in the double pixel, which corresponds to 54\% of the full statistics (including the late signals from the side region).

The simulation values reproduce data trend with the bias voltage for both the estimators. On average, the results obtained in simulation agree within 10\% to the ones measured at the PSI test beam. The difference arises from the limitations in the precise measurement of the time reference resolution and in the accurate determination of all the possible extrinsic effects present in a typical test beam environment. Indeed the presence of residual correlations among the absolute times of arrival of the signals could affect the measured time resolution of the silicon sensor. The experimental setup at PSI does not allow the determination of such correlations from data and, as a consequence, the measurements have potential systematic uncertainties above the quoted statistical uncertainties.
In this regard the test beam comparison discussed in this section constitutes a very good preliminary validation of simulation, whose accurate predictions are also confirmed by using independent experimental measurements in laboratory, with a IR laser and electrons from a $^{90}$Sr radioactive source~\cite{Andreatrento}. These studies are the main subject of a manuscript in preparation.
 
\begin{table}[t]
\centering
\caption{
    The values of the 3D-trench silicon sensor time resolution for different bias voltage from (left) simulation and (right) data. Both $\sigma_{\rm core}$ and $\Delta_{68\%}$, are quoted. For data, $\Delta_{68\%}$ corresponds to the smallest interval around the peak including 54\% of the whole statistics (see text).
    The uncertainty on the time reference has been removed.}
    \label{tab:TimeResolution}
    \vspace{0.3cm}
\resizebox{0.75\columnwidth}{!}{
\begin{tabular}{|c|  cc  |c | c c |}
\hline
 setting & \multicolumn{2}{c|}{Simulation } &
 setting & \multicolumn{2}{c|}{Data  } \\  \hline
 $V_{\rm bias}$ 
 & $\sigma_{\rm core}$   & $\Delta_{68\%}$ & 
 $V_{\rm bias}$
 & $\sigma_{\rm core}$ & $\Delta_{68\%}$ \\
 {[V]}  & [ps] & [ps] & {[V]} & [ps] & [ps]  \\ \hline
$-50$  & $18.9\pm0.2$ & $21.9\pm0.8$ 
& $-50$  & $20.7\pm0.3$ & $24.9\pm1.0$ \\
$-100$ & $16.7\pm0.2$  & $20.1\pm0.7$
&$-110$ & $19.8\pm0.2$ & $23.5\pm0.9$ \\
$-150$ & $16.3\pm0.2$  & $19.7\pm0.7$
& $-140$ & $19.0\pm0.2$ & $23.0\pm0.9$ \\
\hline
\end{tabular}
}

\vspace*{10mm}

\end{table}

\subsection{Results from simulation}

Thanks to the good agreement between data and simulation, several quantities can be studied in simulation to understand the characteristics of the 3D-trench silicon sensor used in charged particle detection.

\subsubsection{Detection efficiency}
The measurements of the detection efficiency as a function of the track impact point position for small pixels requires a sophisticated telescope apparatus with a spatial resolution of a micron level which was not available during the PSI beam test. As a consequence, it is not possible to establish whether the triggered signals in data are representative of the full sensor active area or if they are generated only in a specific portion of it. The simulation, instead, includes the particle impact point information allowing to estimate the double pixel detection efficiency, defined as the fraction of simulated events with signals that exceed the same trigger threshold used at the test beam (see section~\ref{subsec:tfboost_flow}). 

\begin{figure}[t!]
    \centering
    \includegraphics[width=0.45\textwidth]{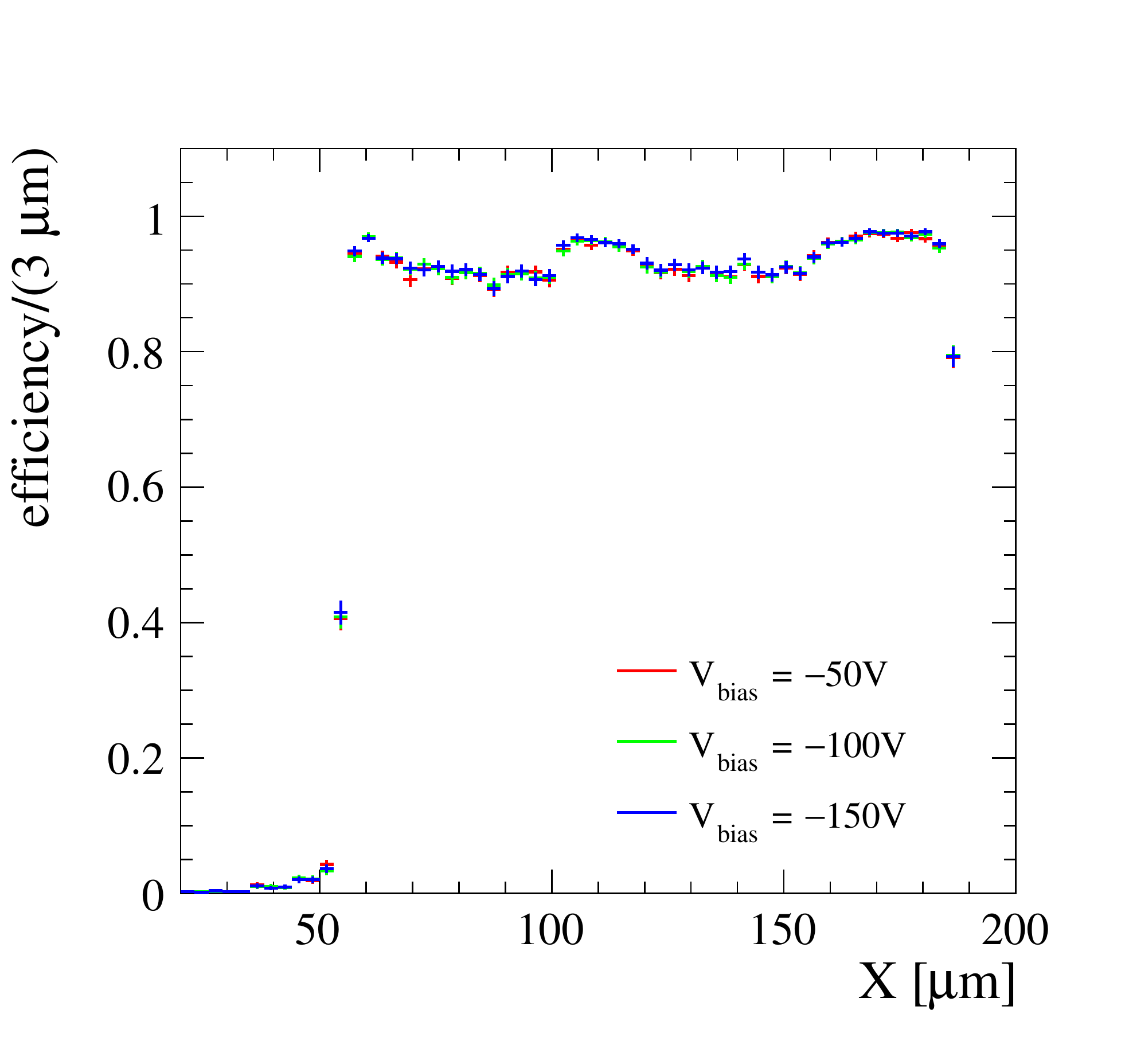}
    \includegraphics[width=0.45\textwidth]{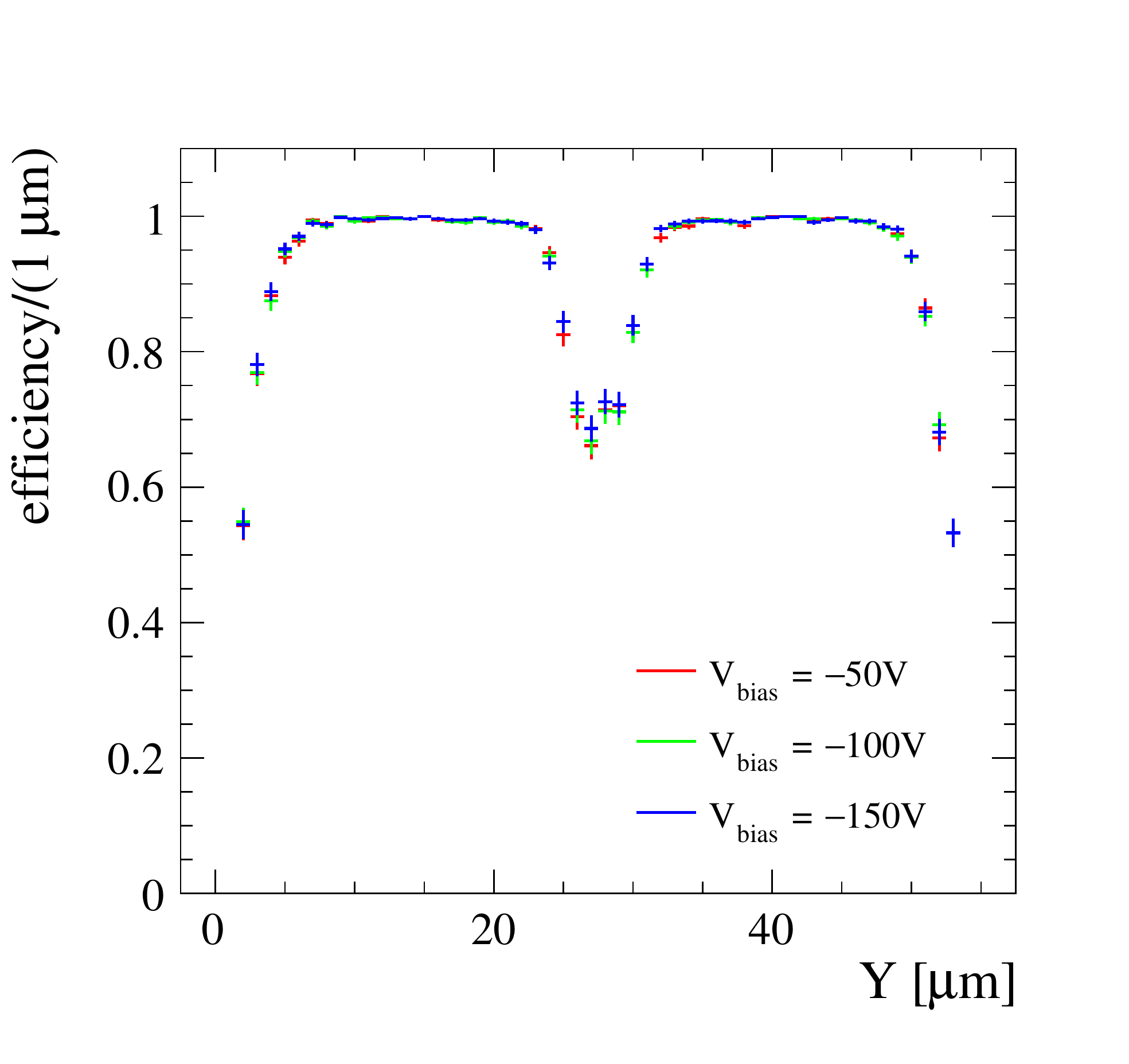}\\
    \caption{Projection over the (left) X and (right) Y coordinate of the track impact point on the surface of the silicon sensor of the detection efficiency. The simulated tracks hit the sensor with an angular distribution similar to data (see section~\ref{sec:mipdeposit}). For the Y projection, only signals with X $>55 \, \mu$m are considered.}
    \label{fig:effXY}
\end{figure}

Figure~\ref{fig:effXY} shows the X and Y projections of the detection efficiency, that is flat in the region of the double pixel ($55 \,\mu${\rm m} <$~$X$ < 165\, \mu$m) and in the active region next to it ($165\, \mu{\rm m} < $X$ < 189\, \mu$m), except for the areas of the double pixel where trenches are located (Y$ <2\, \mu$m, $| {\rm Y}-27.5 |<2\, \mu$m and Y$ >53\, \mu$m). 
Considering that the simulation reproduces the conditions of the test beam (i.e. noise, front-end electronics and trigger threshold), the efficiency is almost maximal for all bias voltages.
%%%%%%%%%%%%%%%%%%%%%%%%%%%
\subsubsection{Timing performances}\label{sec:TimeResolutionSimulationDistr}
\begin{figure}[t!]
\centering
\includegraphics[width=0.65\textwidth]{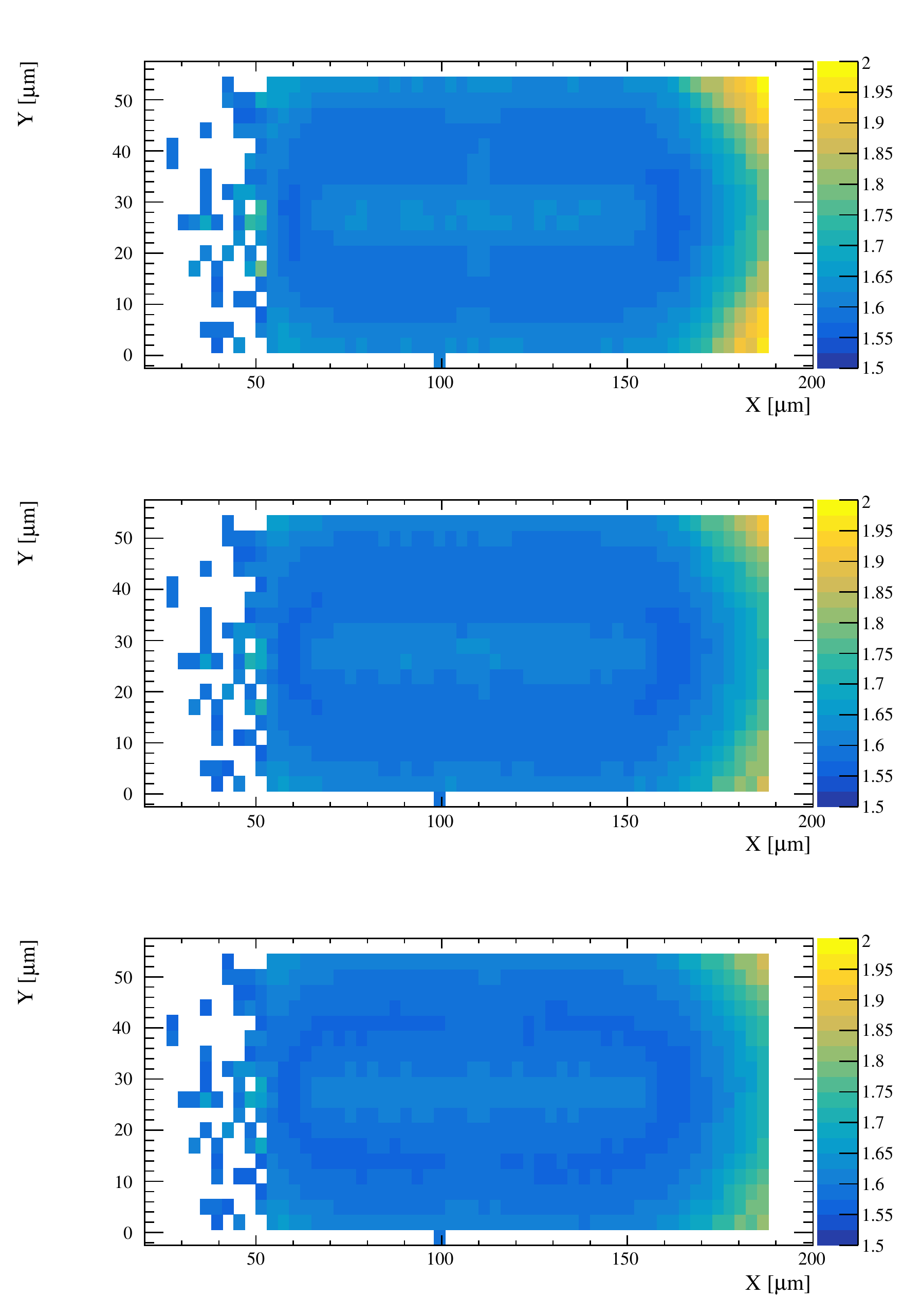}
\caption{Distribution of the signal mean time of arrival with respect to the (X,Y) track impact point coordinates. Plots correspond to simulated samples at (from top to bottom) $V_{\rm bias} = -50, -100$ and $-150$~V.}
\label{fig:timeXY_Vbias}
\end{figure}

%%%%%%%%%%%%%%%%
%%   TIMING   %%
%%%%%%%%%%%%%%%%
The simulation tool allows to investigate in detail the TOA distribution measured at the PSI beam test. The distributions of the signal TOA with respect to the track impact point coordinates are shown in figure~\ref{fig:timeXY_Vbias} for different bias voltages. 

It is clear that a large number of slower events are produced in a small region on the right of the double pixel, $165\, \mu{\rm m} <$~X~$< 189 \,\mu$m (\emph{side region}). In fact in this region the electric field is lower but sufficient to produce a signal that exceeds the threshold. The TOA are typically larger than those from the double-pixel core and vary as much as $200$~ps. 
In a real detector made of a 3D pixel matrix, the contribution of the \emph{side region} might possibly affect only the pixels located at the borders of the matrix. Since this zone is not representative of a double pixel sensor, it must be excluded in the timing characterisation of the double pixel.

\begin{figure}[t!]
    \centering
    \includegraphics[width=0.45\textwidth]{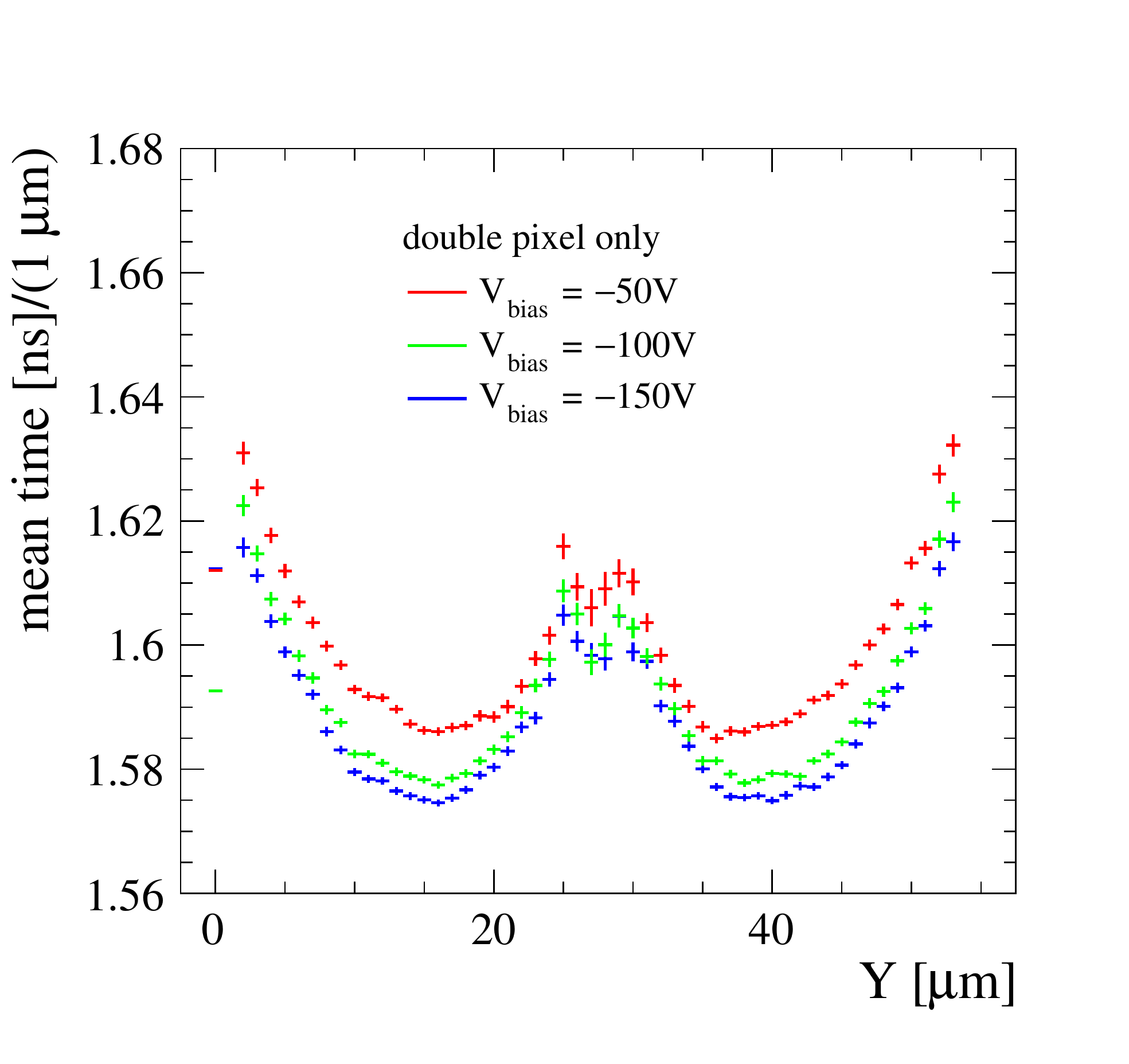}
    \caption{Y projection of the mean time of arrival at different bias voltages. Only signals originated in the double pixel ($55 \, \mu{\rm m} <$~X~$< 165 \, \mu{\rm m}$) are considered.}
    \label{fig:meanTOAprojections_Vbias}
\end{figure}

Focusing on the region of the double pixel, $55 \, \mu{\rm m} <$~X~$< 165 \, \mu$m, the TOA distribution becomes more uniform as the bias increases. The Y projection of the mean TOA, shown in figure~\ref{fig:meanTOAprojections_Vbias}, has a dependency on the distance from the trenches and it changes with the bias voltage.
In particular, the signals originated closer to the bias or the readout trenches are slower than those originated in the middle of the two trenches, because only one of the charge carrier contributes mostly to the current induction, which is then reduced in amplitude and with a longer duration due to the longer drift path. The pattern is not symmetric because of the different velocities of the two charge carriers.

The small region between the two readout trenches of the double pixel core, $105 \, \mu{\rm m} <$~X~$<  115 \, \mu$m and $17\, \mu {\rm m}\le$~Y~$\le 27\, \mu$m) has a smaller electric field. Signals produced in this region are typically slower (and by consequence smaller in amplitude) than those produced in the rest of the double pixel.
The region on the left of the double pixel, X$ < 55 \, \mu$m, is another active pixel, properly biased but not readout. Signals originated there are mostly collected by the corresponding readout electrode and only in a minimal fraction by the double pixel.

\begin{figure}[t!]
    \centering
    \includegraphics[width=0.45\textwidth]{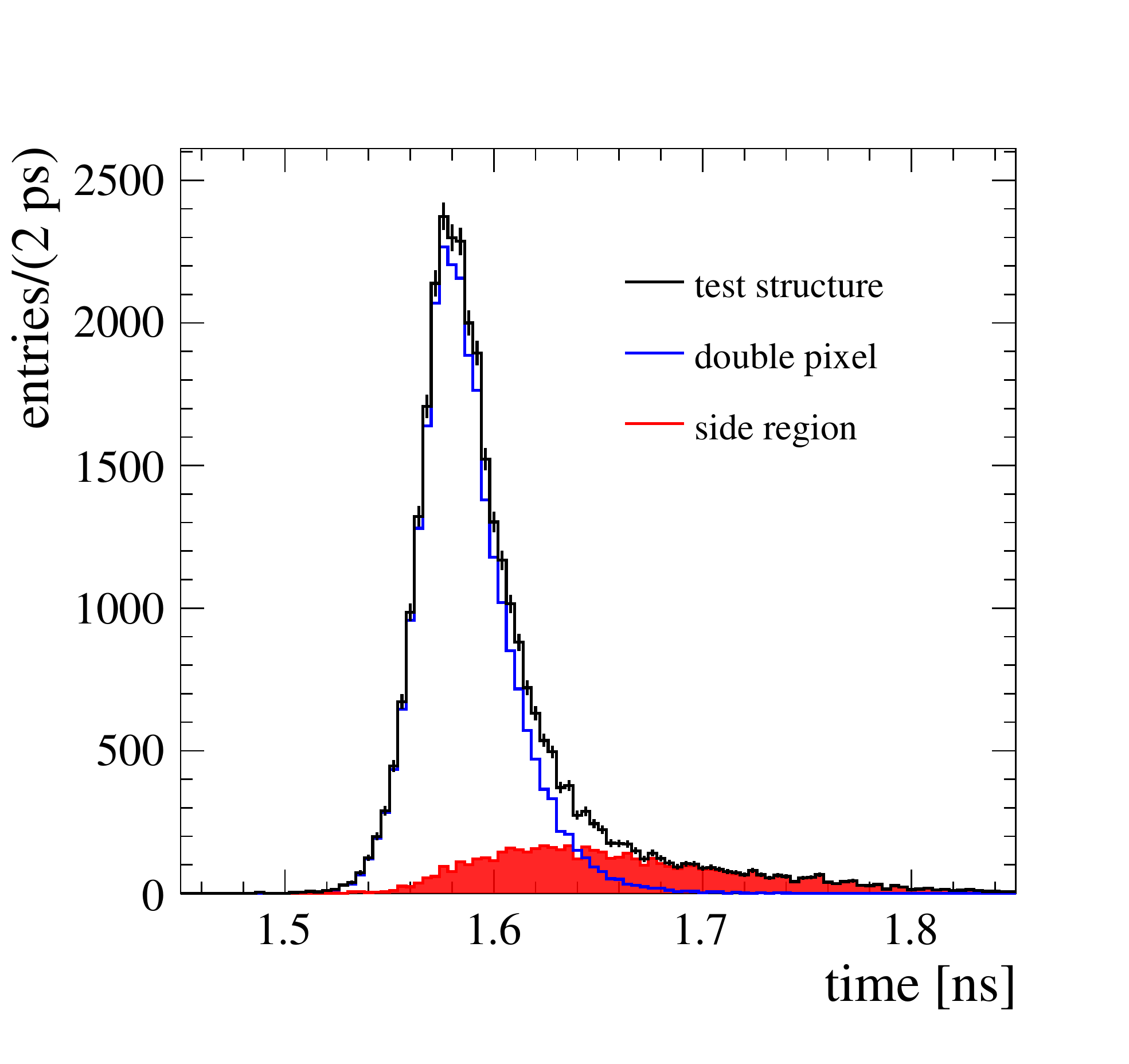}
    \caption{Distributions of the time of arrival for simulated signals at a bias of $-150$~V. All reconstructed signals in the test structure are included, where the contributions due to signals originated in the double pixel and in the low-field side region are overlaid.} 
    \label{fig:SummaryResults_timeResolution}
\end{figure}

The distribution of the TOA of the silicon sensor simulated response at $V_{\rm bias} = -150$~V is shown in figure~\ref{fig:SummaryResults_timeResolution}. The simulation allows to highlight the single contribution from the double pixel region and the side region. The tail is mainly due to signals originated in the side region, while the signals originated in the double pixel contribute to the peaking structure, originating a low asymmetric distribution.

%%%%%%%%%%%%%%%%%%%%%%%%%%%%%
%%  INTRINSIC RESOLUTION   %%
%%%%%%%%%%%%%%%%%%%%%%%%%%%%%

\subsubsection{Intrinsic time resolution of double pixel}\label{sec:IntrinsicTime}
\begin{figure}[tb!]
    \centering
    \includegraphics[width=0.45\textwidth]{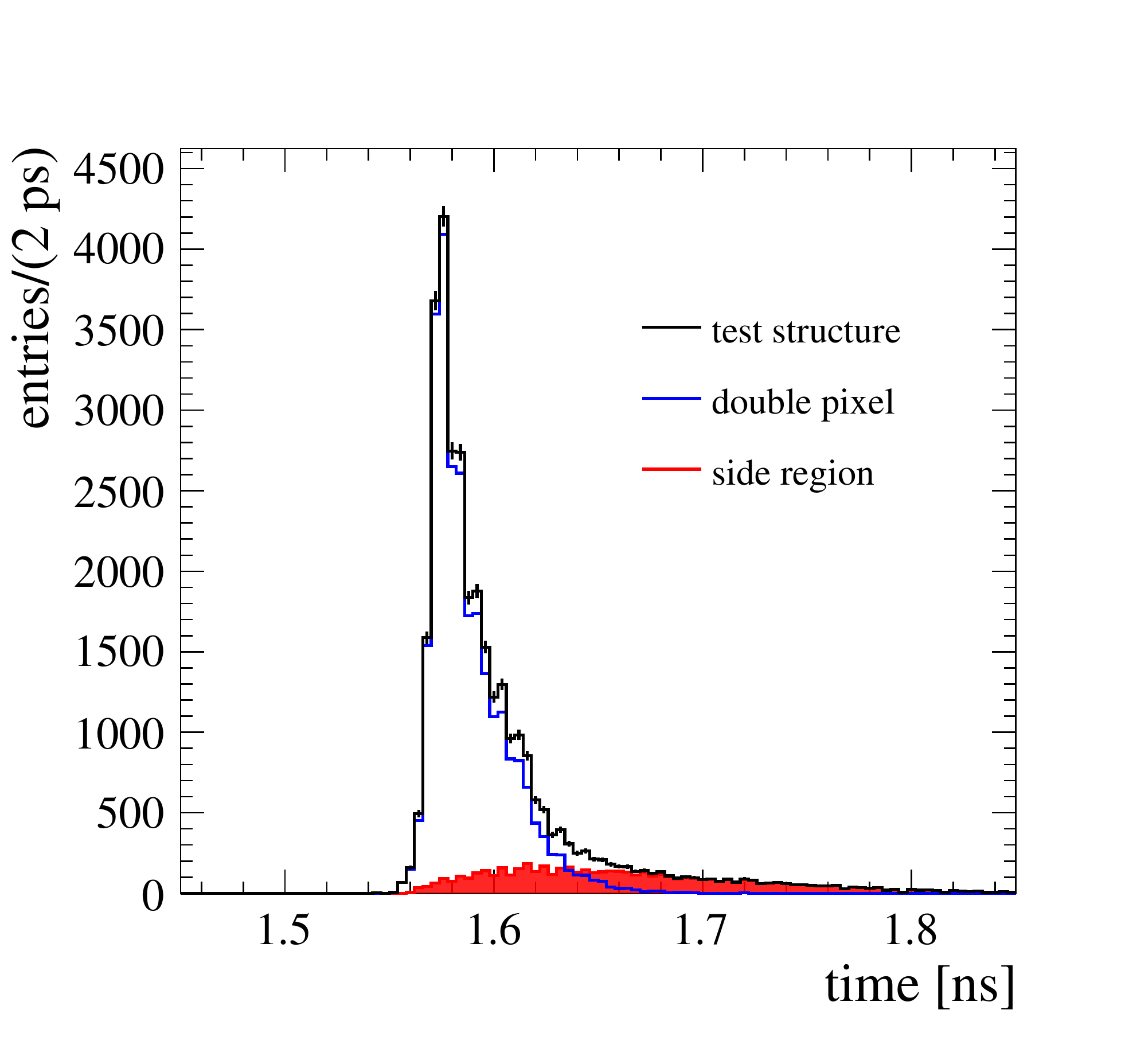}
    \caption{Distributions of the time of arrival for  simulated signals at a bias of $-150$~V without the contribution of the electronic jitter.}
    \label{fig:TimeResolution150_noNoise}
\end{figure}

As reported in ref.~\cite{JINST-TimeSpot}, at a first-order approximation the time resolution of the 3D-trench silicon sensor studied in this paper can be written as
\begin{equation}
 \sigma_t = \sqrt{\sigma_{\rm un}^2 + \sigma_{\rm ej}^2},
 \label{eq:sigma_un}
\end{equation}
where $\sigma_{\rm un}$ is the intrinsic time resolution caused by unevenness in the signal shapes and $\sigma_{\rm ej}$ is the  electronic jitter, which depends on the front-end electronics rise time and signal-to-noise ratio.
The simulation offers the possibility to exactly evaluate the $\sigma_{\rm un}$ contribution, by considering the time of arrival distribution without adding the noise, that is shown in figure~\ref{fig:TimeResolution150_noNoise} for the double pixel at $-150$~V bias. 

The intrinsic TOA distribution of the sensor is asymmetric and shows a narrow peak. The asymmetry is mainly due to the different drift path lengths of charge carriers before being collected by the electrodes, as already shown in figure~\ref{fig:meanTOAprojections_Vbias}, while the long tail comes from the side region of the test structure.

The double pixel intrinsic time resolution is reported in table~\ref{tab:TimeResolution-nonoise}. The $\Delta_{68\%}$ parameter represents the double pixel contribution, while the $\sigma_{\rm core}$, obtained from the fit described in section~\ref{sub:TOA}, represents the contribution of the fastest signals.

\begin{table}[b!]
\centering
    \caption{
    The intrinsic time resolution values ($\sigma_{\rm core}$, $\Delta_{68\%}$) of the double pixel from simulation without the noise contribution.}
    \label{tab:TimeResolution-nonoise}
    \vspace{0.3cm}
\resizebox{0.35\columnwidth}{!}{
\begin{tabular}{|c| c c| }
\hline
 $V_{\rm bias}$ 
 & $\sigma_{\rm core}$  & $\Delta_{68\%}$ \\
 {[V]} & [ps] & [ps] \\ 
\hline
$-50$  &  $9.6\pm0.1$ & $15.4\pm0.7$ \\
$-100$ & $8.0\pm0.1$ & $14.2\pm0.7$ \\
$-150$ & $7.0\pm0.1$ & $14.0\pm0.5$ \\

\hline
\end{tabular}
}
\end{table}

These results are in agreement with those presented in ref.~\cite{JINST-TimeSpot}, where the intrinsic contribution was estimated to be around 15~ps, and show how the 3D approach with an optimised geometry allows to get outstanding time resolution never seen before on silicon sensors without a gain mechanism.

\subsubsection{Electronic jitter contribution}
\begin{figure}[t!]
    \centering
    \includegraphics[width=0.45\textwidth]{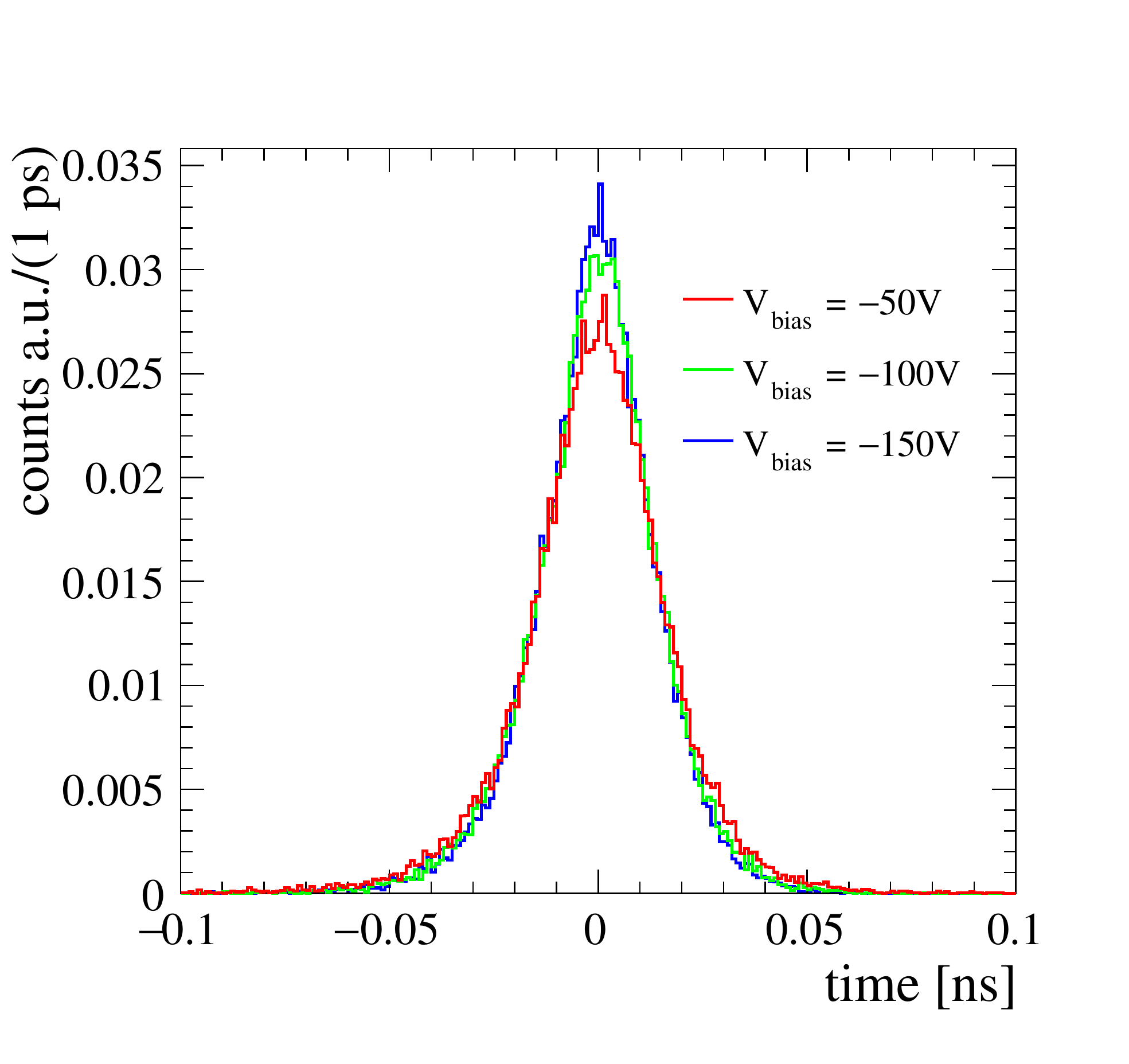}
    \caption{Event-by-event difference between the TOA measured on simulated samples with and without the noise for different values of bias voltage.}
    \label{fig:sigma_ej}
\end{figure}

Similarly to the study discussed in the previous section, the simulation offers the possibility to evaluate the $\sigma_{\rm ej}$ contribution, that can be determined by comparing the TOA measured in the simulation with and without noise. In particular $\sigma_{\rm ej}$ is the width of the distribution of the difference between the TOA in the presence of noise and the TOA without noise, computed for each simulated waveform. The resulting distribution indeed represents the time dispersion introduced by the noise effects only, having the sensor contribution been subtracted event-by-event. The distributions for the three different biases are shown in figure~\ref{fig:sigma_ej} and are characterised by a more evident symmetry with respect to the intrinsic resolution, as expected from a noise induced time dispersion. The tiny residual asymmetry on the left tail is due to the slightly higher probability of finding an earlier TOA due to a fluctuation of the noise. The widths of the distributions range from 18 to 15~ps, with the increasing of the bias voltage absolute value, as a result of the corresponding increasing of the signal dV$/$dt.
Combining this result with the intrinsic time resolution of the double pixel and using the eq.~\ref{eq:sigma_un} the overall time resolution is found to be compatible with the $\Delta_{68\%}$ parameter measured in simulation (table~\ref{tab:TimeResolution}).

\section{Conclusions}
\label{sec:conclusion}
The 3D-trench silicon pixel sensors studied for the TIMESPOT project were designed and produced to address the need to measure with precision the particle arrival time in the next-generation vertex detectors in experiments operating at very high instantaneous luminosities.
By using a beam of minimum ionising particles, the timing performance of a specific test structure read by a custom front-end electronics was measured to be around 20~ps~\cite{JINST-TimeSpot}, which accomplishes the requests of the above mentioned detectors.

The detailed simulation of the tested structure and the comparison of the simulation results with the measurements are presented this paper.
Each step of the simulation, from the modelling of the energy deposit in the sensor by the particle beam to the description of the front-end electronics response, was optimised to guarantee the best approximation to data and the minimum computational time. 
This approach has allowed the production of large samples of simulated signals that are needed for a detailed comparison with data and an accurate study of the simulated sensor response.

The simulation reproduces the data to a good level of approximation in all the measured quantities. The observed differences can be attributed mainly to sources of systematic uncertainties in the test beam measurements and secondarily to the residual limitations of the simulation accuracy.

This study shows the excellent performances of the 3D-trench silicon sensor both in terms of time resolution and in detection efficiency. 
The distribution of the signal time of arrival shows a composite structure related to the specific geometry of the test structure considered, consisting in a biased double pixel located next to a pixel at ground and to an active region subject to a low electric field.
Signals originated in the double pixel contribute to a peaking structure, nearly Gaussian, corresponding to a time resolution of about 20~ps, depending on the bias voltage.
Signals originated in the low-field side region are responsible for the tail of late signals that characterise the observed distribution. 

The analysis of simulated signals without the contribution of the electronic jitter of the front-end circuit used in the measurements with the particle beam indicates that the contribution to the time resolution due to the intrinsic properties of the double pixel sensor is below 15~ps.
As a consequence, the sensor intrinsic performances can be further exploited by means of an improved front-end electronics.

The results obtained and the clear indications about possible performance improvements place 3D-trench silicon sensor at the cutting edge of the development activities for high-resolution timing sensors suitable for charged particle detection. 
In particular, 3D-trench pixel sensors now stand out as a valuable option to be seriously considered in the design of tracking detectors of high-energy physics experiments of the next decades.    

\acknowledgments
{
This work was supported by the Fifth Scientific
Commission (CSN5) of the Italian National
Institute for Nuclear Physics (INFN), within the Project TIMESPOT and by the ATTRACT-EU initiative, INSTANT project.
}

%\begin{thebibliography}{99}

%\end{thebibliography}

\bibliographystyle{unsrt}%ieeetr}  
\bibliography{000AAA-paper} 

\end{document}